\documentclass[sigconf,review=False]{acmart}

\usepackage{multirow}
\usepackage{amssymb}
\usepackage{enumitem}

\newcommand\blfootnote[1]{%
  \begingroup
  \renewcommand\thefootnote{}\footnote{#1}%
  \addtocounter{footnote}{-1}%
  \endgroup
}

\AtBeginDocument{%
  }

\setcopyright{acmlicensed}
\copyrightyear{2024}
\acmYear{2024}
\acmDOI{XXXXXXX.XXXXXXX}

\acmConference[RecSys'24]{RecSys}{October 14--18, 2024}{Bari, Italy}
\acmISBN{978-1-4503-XXXX-X/18/06}

\begin{document}

\title{ReChorus2.0: A Modular and Task-Flexible Recommendation Library}

\author{Jiayu Li*, Hanyu Li*}
\email{{jy-li20,hanyu-li23}@mails.tsinghua.edu.cn}
\affiliation{%
  \department{DCST, Tsinghua University}
  \institution{Quan Cheng Laboratory}
  \city{Beijing}
  \country{China}
}

\author{Zhiyu He}
\email{hezy22@mails.tsinghua.edu.cn}
\affiliation{%
  \institution{DCST, Tsinghua University}
  \city{Beijing}
  \country{China}
}

\author{Weizhi Ma\dag}
\email{mawz@tsinghua.edu.cn}
\affiliation{%
  \institution{AIR, Tsinghua University}
  \city{Beijing}
  \country{China}
}

\author{Peijie Sun}
\email{sun.hfut@gmail.com}
\affiliation{%
  \institution{DCST, Tsinghua University}
  \city{Beijing}
  \country{China}
}

\author{Min Zhang\dag}
\email{z-m@tsinghua.edu.cn}
\affiliation{%
  \department{DCST, Tsinghua University}
  \institution{Quan Cheng Laboratory}
  \city{Beijing}
  \country{China}
}

\author{Shaoping Ma}
\email{msp@tsinghua.edu.cn}
\affiliation{%
  \institution{DCST, Tsinghua University}
  \city{Beijing}
  \country{China}
}


\renewcommand{\shortauthors}{Li et al.}

\begin{abstract}
  With\blfootnote{*: Both authors contribute equally to the paper.\\ \dag: Corresponding author.} the applications of recommendation systems rapidly expanding, an increasing number of studies have focused on every aspect of recommender systems with different data inputs, models, and task settings. 
  Therefore, a flexible library is needed to help researchers implement the experimental strategies they require. 
  Existing open libraries for recommendation scenarios have enabled reproducing various recommendation methods and provided standard implementations. However, these libraries often impose certain restrictions on data and seldom support the same model to perform different tasks and input formats, limiting users from customized explorations.
  To fill the gap, we propose \textbf{ReChorus2.0}, a modular and task-flexible library for recommendation researchers.
  Based on ReChorus, we upgrade the supported input formats, models, and training\&evaluation strategies to help realize more recommendation tasks with more data types. The main contributions of ReChorus2.0 include: 
  (1) Realization of complex and practical tasks, including reranking and CTR prediction tasks;
  (2) Inclusion of various context-aware and rerank recommenders;
  (3) Extension of existing and new models to support different tasks with the same models;
  (4) Support of highly-customized input with impression logs, negative items, or click labels, as well as user, item, and situation contexts.
  To summarize, ReChorus2.0 serves as a comprehensive and flexible library better aligning with the practical problems in the recommendation scenario and catering to more diverse research needs.
  The implementation and detailed tutorials of ReChorus2.0 can be found at \url{https://github.com/THUwangcy/ReChorus}.
\end{abstract}



\maketitle

\section{Introduction}
\label{sec:intro}


Recommendation systems have gained increasing interest from the academic community, with research addressing a wide range of issues throughout the recommendation system workflow. 
These studies, from theoretical analyses to practical applications, have significantly contributed to the development in this domain.

As research advances, various recommendation tasks, experimental settings, and optimization targets emerge.
Recommendation tasks, starting from rating predictions~\cite{koren2009matrix} and top-k recommendations~\cite{rendle2012bpr,sun2023neighborhood}, have extended to recent interests in CTR/CVR prediction~\cite{yang2022click,rendle2010fm,wang2023unbiased} and re-ranking tasks~\cite{pang2020setrank,pei2019PRM}.
Meanwhile, various contexts are considered in recommendation modeling, including users' historical interactions~\cite{kang2018sasrec,hidasi2015gru4rec}, user and item profiles~\cite{guo2017deepfm,cheng2016widedeep}, and the interaction-varying situational context~\cite{li2024sare,lv2023deep}.
Different requirements of candidate set construction also emerged accordingly, such as negative sampling for Top-k recommendation~\cite{rendle2012bpr}, labeled data for CTR prediction, and impression-based logs for reranking~\cite{li2024sare,pang2020setrank}.
Implementing these diverse configurations requires significant time cost. 
A flexible standard framework that can support flexible settings can free researchers from these engineering details and allow them to focus on theoretical and methodological research, while also effectively avoiding implementation bugs.


\begin{table*}[]
\setlength{\abovecaptionskip}{0cm}  
\setlength{\belowcaptionskip}{-0.2cm} 
\caption{Comparisons between ReChorus2.0 and various existing libraries from four aspects: whether they support different task settings, input formats, recommenders, and customized candidate sets. \textit{Cand.} means candidate, \textit{Rec.} means recommendation, \textit{CARS} means context-aware recommender systems, \textit{Pred.} means Prediction, \textit{Imp.} means Impression, and \textit{Situ.} means situational.
}
\label{tab:relatedCompare}
\small
\begin{tabular}{c|cccc|ccc|ccc|cc}
\toprule
\multirow{3}{*}{\textbf{Library}} & \multicolumn{4}{c}{\textbf{Task}} & \multicolumn{3}{c}{\textbf{Input}} & \multicolumn{3}{c}{\textbf{Recommender}} & \multicolumn{2}{c}{\textbf{Customized Cand. Set}$^1$}  \\
\cmidrule(lr){2-13}
 & \multirow{2}{*}{Top-k Rec.} & \multirow{2}{*}{Rerank} & \multicolumn{2}{c}{CARS for} & \multirow{2}{*}{Imp.} & \multicolumn{2}{c|}{Context} & \multirow{2}{*}{Sequential} & \multirow{2}{*}{Rerank} & \multirow{2}{*}{CARS} & \multirow{2}{*}{Training} & \multirow{2}{*}{Evaluation}\\
 &  &  & CTR Pred. & Top-k Rec. &  & U\&I & Situ. &  &  & &   &  \\
 \midrule
LensKit~\cite{ekstrand2020lenskit} & \checkmark &  &   & &  & \checkmark & &  &  &  &  &  \\
Recommenders~\cite{argyriou2020microsoft} & \checkmark &  & \checkmark & \checkmark &  & \checkmark &  & \checkmark &  & \checkmark &  &   \\
Cornac~\cite{salah2020cornac}  & \checkmark  &  & \checkmark  &  &  & \checkmark &  & \checkmark  &  & \checkmark  &  & \\
Elliot~\cite{Vito2021elliot} & \checkmark &  &  \checkmark & \checkmark &  & \checkmark &  & \checkmark &  & \checkmark & \checkmark & \checkmark  \\
DaisyRec~\cite{sun2022daisyrec}& \checkmark &  &  \checkmark & \checkmark &  &  &  &  &  & \checkmark &  &    \\
RecPack~\cite{lien2022RecPack} & \checkmark & &   &  & &  &  &  &  &  &  &  \\
LibRerank~\cite{liu2022neural} & \checkmark & \checkmark  &    & & \checkmark &  &  & \checkmark & \checkmark &  & \checkmark & \checkmark  \\
RecBole~\cite{zhao2022recbole[2.0]} & \checkmark & & \checkmark &  &  & \checkmark & \checkmark & \checkmark &  & \checkmark &  &   \\
BARS~\cite{zhu2022bars}& \checkmark &  & \checkmark & &  & \checkmark & \checkmark & \checkmark &  & \checkmark &  &     \\
ClayRS~\cite{lops2023clayrs}& \checkmark &  & \checkmark & &  & \checkmark &  &  &  & \checkmark &  &    \\
ReChorus1.0~\cite{wang2020make} & \checkmark &  &  & &  & \checkmark &  & \checkmark & & &  &     \\
\midrule
\textbf{ReChorus2.0} & \checkmark & \checkmark & \checkmark & \checkmark & \checkmark & \checkmark & \checkmark & \checkmark & \checkmark & \checkmark & \checkmark & \checkmark  \\
 \bottomrule
\end{tabular}
\begin{minipage}{18cm}
\vspace{0.1cm}
\footnotesize  
$^1$: \textit{Customized candidate set} indicates that users can customize positive and negative samples with variable lengths in each candidate set.

\end{minipage}

\end{table*}

Recently, there have been a number of excellent libraries to facilitate standard implementations of numerous recommendation algorithms and tasks.
These libraries significantly contribute to the reproducibility of the recommendation research community.
For instance,
BARS~\cite{zhu2022bars}, including BART-CTR and BARS-Batch, targets open benchmarks for both CTR prediction and candidate item matching tasks on a variety of public datasets.
RecBole~\cite{zhao2022recbole[2.0]} provides a unified and comprehensive recommendation framework supporting 43 datasets and 91 algorithms.
There are also other famous libraries, such as the venerable LensKit~\cite{ekstrand2020lenskit} and Microsoft Recommenders~\cite{argyriou2020microsoft}, Cornac for multimodal recommendations~\cite{salah2020cornac}, and LibRerank for reranking methods~\cite{liu2022neural}.
However, as shown in Table~\ref{tab:relatedCompare}, they mostly impose constraints on input data and candidate set construction, and supported tasks are also limited.
This lack of flexibility in altering model and data settings makes it less convenient for researchers to conduct experiments, especially exploratory studies, during research progress.

To tackle these challenges, we provide \textit{\textbf{ReChorus2.0}}, \textbf{a modular and task-flexible library for research in recommendation}.
The updated library is built on ReChorus1.0~\cite{wang2020make}, a swift and efficient framework focusing on sequential recommendations, one of the frameworks recommended by ACM  RecSys2024\footnote{\url{https://github.com/ACMRecSys/recsys-evaluation-frameworks}}.
Leveraging and enhancing the modular nature of ReChorus, we extend it to encompass more complex recommendation tasks in practice, including CTR prediction tasks with various context metadata and ranking/reranking with customized candidate sets for training and evaluations~(i.e., users can customize positive and negative samples with variable lengths in each candidate set such as impression logs). 
By enabling the free assembly of modules akin to building blocks, ReChorus2.0 allows one recommender to support diverse tasks and input formats, helping users implement desired tasks.

To be specific, in this upgrade to ReChorus2.0, several major modifications are implemented: 
(1) The realization of more complex tasks closely aligned with practice, such as reranking and CTR prediction tasks;
(2) To accomplish the extended tasks, more flexible training and evaluation strategies are implemented, including training and evaluations on highly customized candidate sets with variable length and multiple positive instances, as well as binary label classification for CTR;
(3) We expand existing recommenders and introduced new ones, making a single model able to support multiple tasks;
(4) Various input formats are accommodated, including dense and discrete metadata, interaction-varying situation contexts, and impression logs.

ReChorus2.0 marks a substantial update toward a more modular and task-flexible framework. It better aligns with the practical recommendation systems, catering to more diverse research needs. 
We hope ReChorus2.0 will serve as a highly customized and user-friendly tool for more researchers to form a ``Chorus'' of recommendation tasks and algorithms.



%


\section{Related Work} 
\label{sec:relatedWork}

\subsection{Reranking in Recommendations}
\label{ssec:relatedImpression}

Serving as the last step of a recommender system, the re-ranking stage obtains and permute top items from the ranking stage as candidates.
The importance of re-ranking models has gained increasing attention~\cite{liu2022neural}, especially in the industry, since its output directly affects the revenue. 
Even ranking algorithms~\cite{Ren2023SlateAwareRF} begin to implicitly consider the relations among the items to provide high-quality candidate sets for re-ranking. 

Re-ranking algorithms consider the inter-dependencies among items in a candidate list, which can be categorized as two groups: two-step models~(evaluator and generator) and unified models~(maintaining permutation invariance in the candidate list).
Two-step models typically utilize reinforcement learning techniques to maximize the utility. For instance, Aliexpress~\cite{huzhang_aliexpress_2020} employs an LSTM-based evaluator to assess the utility of re-ranked item lists, guiding the generator to determine the optimal permutation of candidate items. 
Other approaches~\cite{feng_grn_2021,xi_context-aware_2022} also focus on maximizing list-wise utility through similar methods.
Unified re-ranking models treat input items as a set, ensuring the permutation of items does not affect the order of re-ranked results. 
PRM~\cite{pei2019PRM} was the first to apply transformer blocks in reranking to model global relationships in the candidate list. 
Setrank~\cite{pang2020setrank} introduced the concept of permutation invariance and proposed to use induced self-attention to extract item correlations. 
And MIR~\cite{xi_multi-level_2022} extracted fine-grained interactions between candidate items and historical interactions. 

Given the importance of reranking tasks in practical applications and the recent trend towards unified reranking models, we support unified reranking models and corresponding training\&evaluations on customized candidate sets in ReChorus2.0.

\subsection{Context-aware Recommendation}
\label{ssec:relatedContext}

Context-aware recommendations consider contextual attributes to capture user preferences more accurately~\cite{kulkarni2020context}.
The broad definition of context, which is adopted in this paper, encompasses user profiles, item content, and interaction-varying situations.
Early works on context-aware RecSys explored various structures to model the relations of context features, such as interaction terms in FM~\cite{rendle2010fm}, a combination of FM and deep neural networks in DeepFM~\cite{guo2017deepfm} and xDeepFM~\cite{lian2018xdeepfm}, and attention-based pooling in AFM~\cite{xiao2017afm}.
More recent models, such as FinalMLP~\cite{mao2023finalmlp} and SAM~\cite{cheng2021sam}, are still exploring better structures to capture relations between features.
As RecSys becomes increasingly complex, users' historical interactions are also considered in context modeling.
For instance, DIN~\cite{zhou2018din} utilized an activation unit to learn the relationships between candidate items\&context and user behavior history. DIEN~\cite{zhou2019dien} adopted an interest-evolving layer to capture user interest from history. CAN~\cite{bian2022can} enhanced sequential recommenders with Co-Action Units between target items, context, and history features.
Recently, situations, i.e., context varying with user interactions, have attracted attention~\cite {lv2023deep,li2024sare}.
In these works, users' location and time are considered for dynamic user preference modeling.

In summary, context-aware recommendation is an evolving research hotspot. The majority of existing work has been conducted under the CTR prediction task~\cite{rendle2010fm,song2019autoint,cheng2021sam,mao2023finalmlp,zhou2018din}, while some recent exploration~\cite{li2024sare,bian2022can} has also applied context information for ranking tasks.
Therefore, in ReChorus2.0, we implement various context-aware models and set them to support both CTR prediction and top-k recommendation tasks. Additionally, we expand the reader to accommodate different formats of context information.


\subsection{Libraries for Recommendation Algorithms}
\label{ssec:relatedLibrary}







The increasing complexity and diversity of recommendation systems have raised concerns about reproducibility and fair comparison in recommendation research~\cite{ferrari2021troubling,sun2020we,sun2023challenges}. 
To address the issues, researchers have made great efforts to design tools and frameworks that promote reproducible research.

Table~\ref{tab:relatedCompare} compares our proposed ReChorus2.0 framework with various existing libraries. While most existing frameworks provide support for basic Top-k recommendation and CTR prediction tasks, few fully support context information for different tasks. For instance, only three of the frameworks support context-aware models for Top-k recommendation task. 
Furthermore, most context-aware frameworks focus on the side information from users and items while neglecting the situational context that varies with interactions. Moreover, ReChorus2.0 can also handle different interaction formats. Besides the ordinary click and rating feedback, the impression logs (exposure but non-click) are also considered.

ReChorus2.0 also distinguishes itself by fully adopting the re-ranking task and corresponding customized candidate sets, which has gained increasing attention in both academics and industry. 
Only Librerank~\cite{liu2022neural} provides a comprehensive re-ranking algorithm platform, but it only offers two fixed base-ranker options, which limits comparing ranking and re-ranking models based on the same evaluation system. 
ReChorus2.0 enables the re-ranking setting by inducing the customized candidate set. 
Apart from the conventional negative sampling operation for training and evaluation, researchers can also customize the positive and negative samples with variable lengths for each candidate set, both in the training and evaluation datasets. 
Moreover, in ReChorus2.0, ranking and re-ranking models can be flexibly combined as they would be in real-world scenarios.

Note that we only compare with libraries that provide generic recommenders aiming at performance improvement. Frameworks with other targets, experiment settings, or specific scenarios are excluded in our discussions, such as reinforcement learning-based recommender framework EasyRL4Rec~\cite{yu2024easyrl4rec}, fairness-aware component in Recbole2.0~\cite{zhao2022recbole[2.0]}, recommender evaluation library RecList~\cite{chia2022RecList}, and news recommendation library~\cite{iana2023newsreclib}.



\section{The ReChorus2.0 Framework}
\label{sec:framework}

\begin{figure*}
    \setlength{\abovecaptionskip}{0cm}  
    \setlength{\belowcaptionskip}{-0.2cm} 
    \centering
    \includegraphics[width=\linewidth]{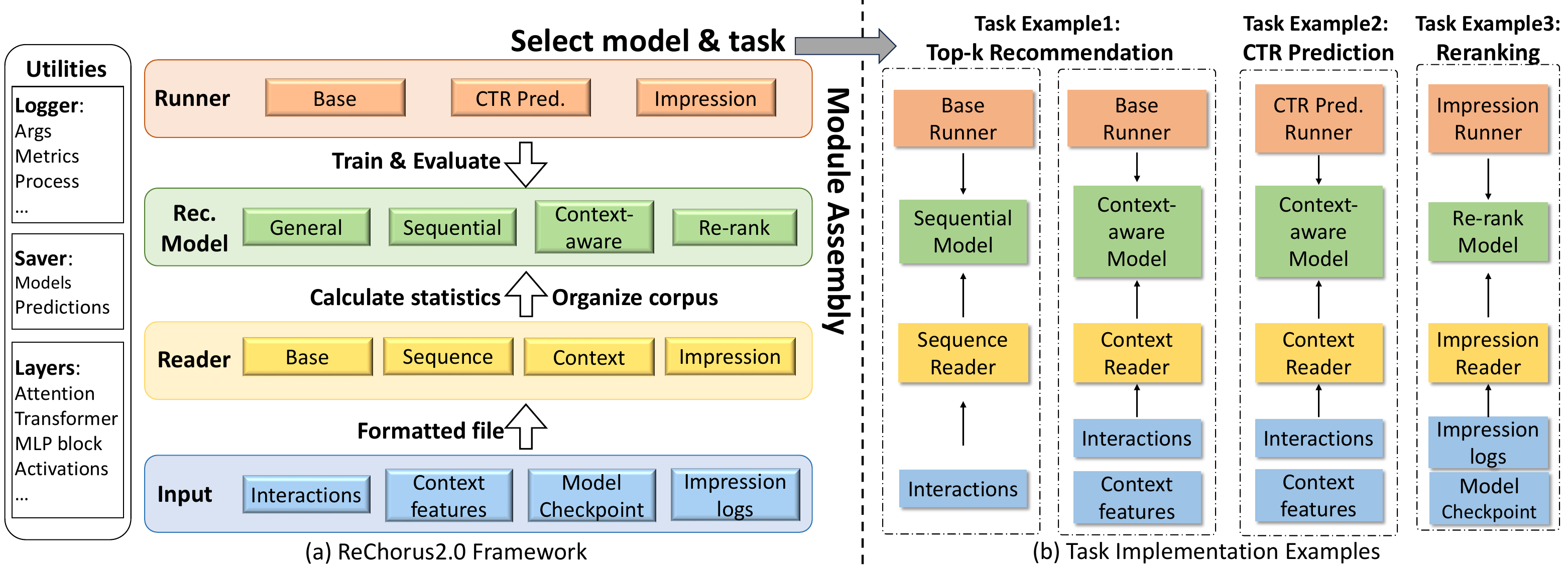}
    \caption{The overall framework of ReChorus~2.0~(a) and some task implementation examples~(b). ReChorus2.0 can fit various recommendation tasks by flexible assembly of four parts: input data, reader, model, and runner. Corresponding modules are automatically assembled after users specify the model and task.}
    \label{fig:Framework}
\end{figure*}

\subsection{Supported Task Categories}
\label{ssec:frameworkTask}


As a task-flexible library, ReChorus2.0 is able to support a variety of tasks with different types of input, models, and training\&evaluation strategies.
We introduce the tasks in general in this section.

\begin{itemize}[nolistsep, leftmargin=*]
\item \textbf{Top-k Recommendation Task}: Top-k recommendation is a common task where the goal is to suggest a list of $k$ items that they are most likely to be interested in for users~\cite{rendle2012bpr,wang2020make}.
During both training and evaluation, a fixed number of randomly sampled unseen items are treated as the negative items for each positive interaction.
List-wise evaluation metrics, such as Hit Ratio and NDCG, are calculated on lists containing one positive item and a fixed number of negative items.
In ReChorus2.0, general, sequential, and context-aware recommenders all support the Top-k recommendation task.
\item \textbf{Impression-based Ranking/Reranking Task}: Different from top-k recommendation, impression-based tasks do not involve negative sampling. Instead, the candidate item lists shown to the user (i.e., the impression item list) are used, and the non-clicked items are viewed as negative in each impression. 
In practice, impression lists may be variable in length, and the number of positive and negative items in one impression list also varies.
The evaluation metrics are calculated over all positive interactions in the same candidate list before being averaged among lists.
In ReChorus2.0, general, sequential, and reranking models all support the impression-based task.
\item \textbf{Click-Through Rate~(CTR) Prediction Task}: Click-Through Rate (CTR) prediction aims to estimate the probability that a user will click on a given item, and the estimation is compared with a binary label representing click/non-click. 
Evaluation metrics, such as LogLoss and AUC, are calculated based on the prediction probabilities and ground truth labels of all interactions or each user.
Generally, rich context information is utilized for CTR prediction tasks.
In ReChorus2.0, context-aware models can support this task, where user profiles, item metadata, and situation context~(i.e., context varying with interactions) can be utilized.
\end{itemize}

\subsection{Overall Framework Design}
\label{ssec:frameworkOverall}
As shown in Figure~\ref{fig:Framework}(a), ReChorus2.0 follows the flexible modular design and introduces numerous new features within each part of module to support a wider variety of models and task settings.
In general, Rechorus2.0 supports various recommendation tasks from different \textbf{Input} formats by assembly of three main modules: \textbf{Reader}, \textbf{Model}, and \textbf{Runner}.
\textbf{Reader}s deal with inputs, calculate the dataset statistics, and organize data as a unified corpus for models. Basic interactions, click labels, impressions, model checkpoints, and abundant context information are all acceptable inputs.
Diverse recommendation \textbf{Model}s are supported, including generic and sequential models, context-aware recommender systems~(CARS), and re-rank recommenders.
Up to 37 recommendation methods are supported in ReChorus2.0.
Depending on the data and the specific task, the corresponding \textbf{Runner} is invoked: Base Runner for top-k recommendation, CTR Runner for CTR prediction, and Impression Runner for impression-based ranking \& re-ranking tasks.
Note that readers and runners are automatically assembled to models once users select the recommendation model and mode~(i.e., task), freeing users from implementation details and allowing them to focus on methodological and experimental exploration.
Specifically, functions about the context-aware CTR predictions and recommendations, as well as impression-based (re-)~ranking are all newly introduced in ReChorus2.0.

Besides the main modules, utility modules are adopted to back all modules, including a logger to write the running processes, a saver for model checkpoints and prediction, and several common neural network layers.
Moreover, an experiment script, \textit{exp.py}, is designed to automatically conduct repeat experiments with different random seeds and save the average performances.

\subsection{Data Readers}
\label{ssec:frameworkReader}

\begin{figure}
\setlength{\abovecaptionskip}{0cm}  
\setlength{\belowcaptionskip}{-0.2cm} 
    \centering
    \includegraphics[width=\linewidth]{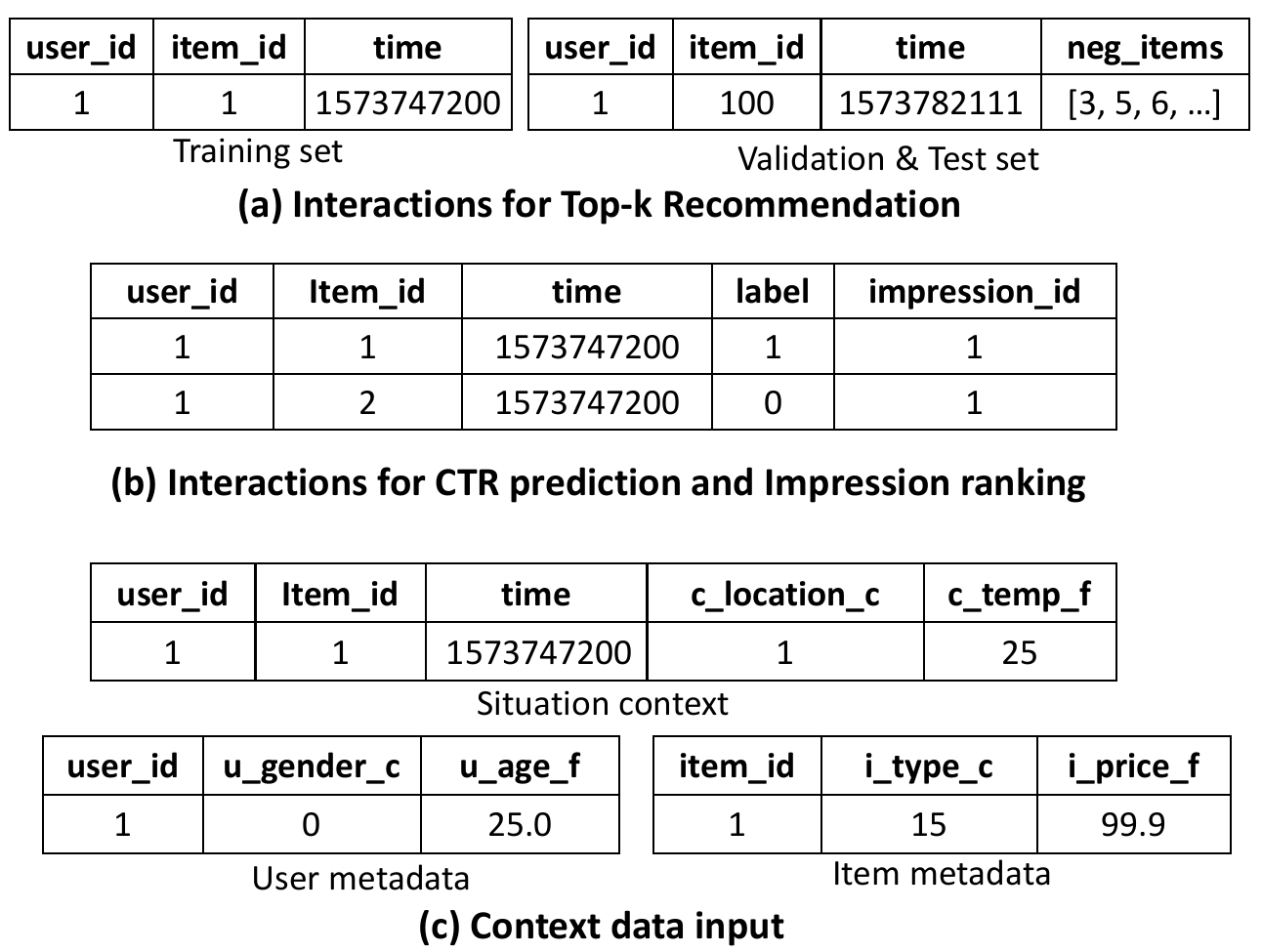}
    \caption{Examples of data input formats supported by ReChorus2.0. Note that \textit{impression\_id} is only necessary when conducting impression-based tasks in Figure(b).}
    \label{fig:dataFormat}
\end{figure}

\textit{Reader} modules are utilized to transfer input files to a unified corpus for training and evaluation.
During the transfer, basic dataset statistics are recorded. 
After processing, the entire corpus is stored as a \textit{.pkl} file to facilitate direct loading in future uses, thereby saving time.
Various types of readers are designed to support flexible inputs with different formats and content, as shown in Figure~\ref{fig:dataFormat}:
\begin{itemize}[nolistsep, leftmargin=*]
    \item \textbf{Base reader}: It is the basic class for all readers and is directly called by Top-k recommendation tasks. Base readers support input of interactions with user and item IDs, with negative items assigned to validation and test sets, as shown in Figure~\ref{fig:dataFormat}(a).
    \item \textbf{Sequential reader}: Built on the base reader, sequential readers capture and save users' interaction history for sequential recommender systems to support Top-k recommendation tasks.
    \item \textbf{Impression reader}: It enables impression data reading of the format in Figure~\ref{fig:dataFormat}(b). Interactions with the same $impression\_id$ are grouped, and later, they will be trained and evaluated together. In this way, users can customize their desired ways to construct any type of candidate set.
    \item \textbf{Context reader}: It helps read contextual information of users, items, and interaction situations, as shown in Figure~\ref{fig:dataFormat}(c). 
\end{itemize}

Besides these four fundamental readers, ReChorus2.0 provides their combinations for various data format requirements, such as \textit{Context-SeqReader} for context-aware sequential recommenders and \textit{Context-ImpReader} for context-aware impression-based tasks.



It is worth noting that, unlike some popular frameworks~\cite{zhao2022recbole[2.0],zhu2022bars}, ReChorus does not provide dataset filtering and split processes.
On the contrary, training, validation, and test sets are fed into the readers separately, with post-processing to combine them if necessary~(e.g., for user history or impression construction).
This allows users to freely customize dataset settings, facilitating deep investigation of datasets in academic research.

\subsection{Supported Models}
\label{ssec:frameworkModel}

In this upgrade of ReChorus, we add numerous context-aware and re-ranking recommenders and expand existing models to cover more tasks, adding up to 37 recommenders. In the implementation, we enable many models to effectively support more than one task by model inheritance and code reuse.
Users can switch to different tasks easily by changing the ${model\_mode}$ configuration.
Models and their supported modes~(i.e., tasks) are listed as follows:
\begin{itemize}[nolistsep, leftmargin=*]
    \item \textbf{General models}: This model group includes CF-based recommenders considering user and item IDs, which support \textit{Top-k recommendation} and \textit{impression-based ranking} tasks, including:
    Most popular, BPR~\cite{rendle2012bpr}, Neural MF~\cite{he2017neuCF},
    CFKG~\cite{zhang2018cfkg}, LightGCN~\cite{he2020lightgcn}, BUIR~\cite{lee2021buir},  and DirectAU~\cite{wang2022directau}.
    \item \textbf{Sequential models}: Sequential models utilize users' historical item interaction sequences to recommend users' next interaction, which support \textit{Top-k recommendation} and \textit{impression-based ranking} tasks, including: FPMC~\cite{rendle2010fpmc}, 
    GRU4Rec~\cite{hidasi2015gru4rec}, NARM~\cite{li2017narm},
    Caser~\cite{tang2018caser}, SASRec~\cite{kang2018sasrec}, SLRCPlus~\cite{wang2019SLRCPlus}, Chorus~\cite{wang2020make}, ComiRec~\cite{cen2020comirec}, KDA~\cite{wang2020kda}, TiSASRec~\cite{li2020tisasrec},       TiMiRec~\cite{wang2022timirec}, and ContraRec~\cite{wang2023contrarec}.
    \item \textbf{Re-rank models}:
    Re-rank models incorporate pre-trained backbone recommenders to provide ranking lists and re-sort them. Both general and sequential backbones are supported for \textit{impression-based ranking} with PRM~\cite{pei2019PRM}, SetRank~\cite{pang2020setrank}, and MIR~\cite{xi_multi-level_2022}.
    \item \textbf{Context-aware models}: 
    Context-aware recommender systems (CARS) adopt diverse structures to capture the influences of contextual information on users' preferences.
    Both interaction-based and sequential CARS support \textit{Top-k recommendation} and \textit{CTR prediction} tasks, including:
    FM~\cite{rendle2010fm}, Wide\&Deep~\cite{cheng2016widedeep}, DeepFM~\cite{guo2017deepfm}, AFM~\cite{xiao2017afm}, DCN~\cite{wang2017dcn}, xDeepFM~\cite{lian2018xdeepfm},
    AutoInt~\cite{song2019autoint},  DCNv2~\cite{wang2021dcnv2}, FinalMLP~\cite{mao2023finalmlp}, SAM~\cite{cheng2021sam}, SARE~\cite{li2024sare},
    DIN~\cite{zhou2018din}, DIEN~\cite{zhou2019dien}, CAN~\cite{bian2022can} (based on DIEN), ETA~\cite{chen2021eta}, and SDIM~\cite{cao2022sdim}.
\end{itemize}




It is important to note that since ReChorus2.0 targets at offering a modular and flexible framework, we only include classic and some state-of-the-art recommenders in each model group.
Owing to the extensibility of ReChorus 2.0, users can easily implement new models~(Details are shown in Section~\ref{ssec:usageExtension}). 
We are also planning to add more models in the future.

\subsection{Training and Evaluation Runners}
\label{ssec:frameworkRunner}

Runners are responsible for training and evaluating the models based on a specified task, providing appropriate training strategies and evaluation metrics.
\begin{itemize}[nolistsep, leftmargin=*]
\item \textbf{Base Runner}: The base runner is adopted for Top-k recommendation tasks, where BPR losses are used for optimization, and Hit Rate~(HR) and NDCG~\cite{jarvelin2002cumulated} are used as metrics.
\item \textbf{Impression Runner}: It contributes to the variable-length and multiple-positive recommendation tasks, where multiple loss functions can be adopted, including list-level BPR loss~\cite{li2023intent}, listnet loss~\cite{Cao2007LearningTR}, softmax cross-entropy loss, and attention rank~\cite{ai_learning_2018}. Evaluation metrics include HR, NDCG, and MAP, where we adopt matrix operations to optimize the time efficiency of calculations.
\item \textbf{CTR Runner}: CTR runner is designed for CTR prediction tasks, with BPR/BCE loss for training and AUC/Log loss for evaluation.
\end{itemize}

For expression consistency, loss functions are introduced here, along with the runners. 
In the implementation, loss functions are actually implemented in the task-oriented \textit{base model}s. By inheriting different base models in different modes, each recommender, as introduced in the previous section, can be applied to varied tasks.

\begin{figure*}[htbp]
\setlength{\abovecaptionskip}{0cm}  
\setlength{\belowcaptionskip}{-0.2cm} 
    \includegraphics[width=\linewidth]{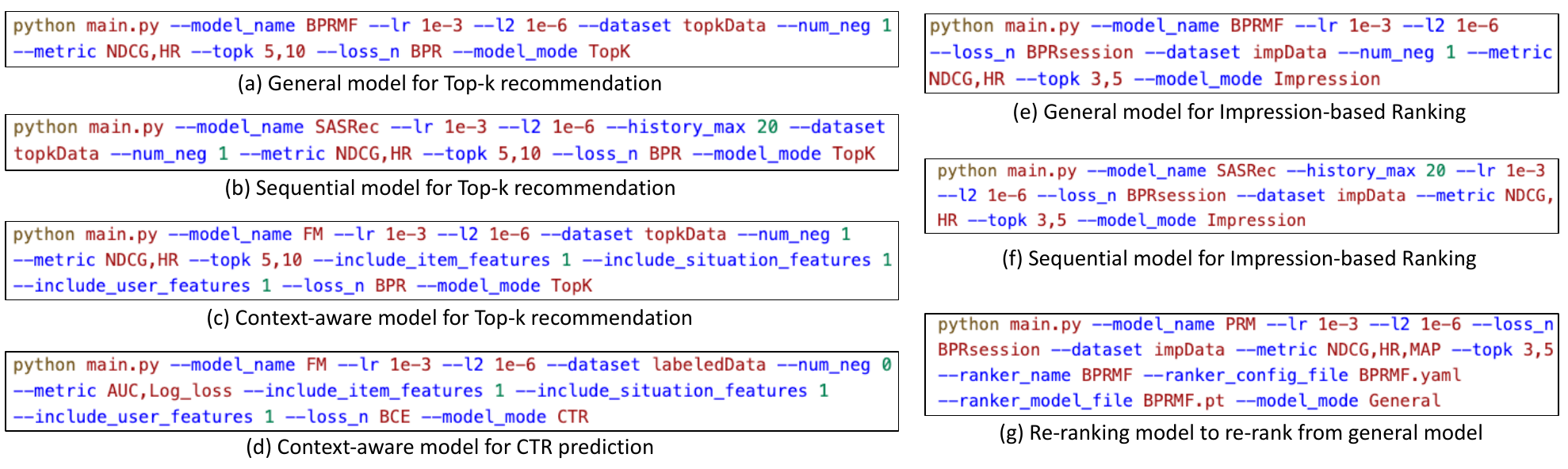}
    \caption{Examples of configurations to run different tasks with various types of recommenders.}
    \label{fig:configuration}
\end{figure*}

\section{Task-based Usage Guidelines}
\label{sec:usage}



Given that the update of ReChorus2.0 is centered on task enhancement, we present several examples to demonstrate the flexible implementation of various tasks using ReChorus2.0 as user guidelines.
Examples of configurations to run all supported task categories~(in Section~\ref{ssec:frameworkTask}) are shown in Figure~\ref{fig:configuration}, and details about usage tutorials for all these existing and new tasks can be found in the repository\footnote{\url{https://github.com/THUwangcy/ReChorus/tree/master/docs/tutorials}}.
In the following subsections, we will focus on three main task categories newly supported in ReChorus2.0 and illustrate the data preparation and configuration settings for each task: Impression-based Ranking/Re-ranking in Figure~\ref{fig:configuration}(e)-(g), CTR prediction in Figure~\ref{fig:configuration}(d), and context-aware Top-k recommendation in Figure~\ref{fig:configuration}(c). 
Moreover, extensions to new tasks and models will also be briefly introduced.



\subsection{Impression-based Ranking/Re-ranking}
\label{ssec:usageImpression}

In impression-based ranking and re-ranking tasks, each sample's candidate set may vary in length and contain arbitrary number of positive and negative instance. 
To standardize processing in the library, ReChorus2.0 requires that interaction data be formatted with each interaction on one line, identified by an \textit{impression ID} to denote interactions under the same impression, as shown in Figure~\ref{fig:dataFormat}(b). The \textit{impression reader} will automatically aggregate multiple interactions sharing the same impression ID for training and testing.

Examples of running configurations are illustrated in Figure 3(e)-(g). For ranking tasks, users simply need to set $model\_mode$ to 'Impression'. For re-ranking tasks, it is also necessary to specify the name, configuration, and checkpoint of the backbone ranker to obtain ranking inferences from the backbone.


\subsection{CTR Prediction Task}
\label{ssec:usageCTR}

As defined in Section~\ref{ssec:frameworkTask}, 
the CTR prediction task aims to predict whether a user will click on a given item, essentially a binary classification problem. 
To undertake a CTR prediction task with ReChorus 2.0, a dataset comprising interactions and corresponding labels~(click or not) should be prepared, as depicted in Figure~\ref{fig:dataFormat}(b)~(Note that $impression\_id$ is not necessary). 
Furthermore, users can input context by appending situation information in the interaction file, as well as extra $user\_metadata.csv$ and $item\_metadata.csv$ files, as shown in Figure~\ref{fig:dataFormat}(c).

Once the dataset is prepared, users simply need to specify the desired recommendation model and set the $model\_mode$ parameter to \textit{CTR} in the execution command. An example of the main parameters is illustrated in Figure~\ref{fig:configuration}(d), where the $include\_XXX\_features$ parameter allows for flexible input of partial or all prepared context, while the $num\_neg$ parameter should be set to 0, since negative sampling is unnecessary for CTR prediction tasks.


\subsection{Context-aware Top-k Recommendation Task}
\label{ssec:usageTopk}

Top-k recommendation, where recommender models rank either the entire item set or a randomly sampled subset of items and provide Top-k items as recommendation results, was the only task supported by ReChorus1.0. 
In ReChorus2.0, we enable the newly added context-aware models to support the Top-k recommendation task as well.
The dataset preparation for this task includes a training set containing interactions and validation\&test sets containing interactions and corresponding negative item candidates, as shown in Figure~\ref{fig:dataFormat}(a) (for full-set evaluation, $neg\_items$ is not needed). Additionally, context information, as depicted in Figure~\ref{fig:dataFormat}(c), is also permitted as input.

An example for task running configurations is shown in Figure~\ref{fig:configuration}(c), where $model\_mode$ should be set to \textit{TopK} and $num\_neg$ should be greater than zero for training with negative sampling and the BPR loss function~\cite{rendle2012bpr}.



\subsection{Extension to New Models and Tasks}
\label{ssec:usageExtension}

Since ReChorus2.0 follows the flexible modular design, users can conveniently extend to new models and tasks in ReChorus2.0.

To implement a new model for an existing task, users simply need to create a new model class, inherit the corresponding base model, and specify the \textit{runner} name corresponding to the task and the \textit{reader} name matching the data. 
Within the new model class, users only need to define the model's required parameters and construct a forward function that accepts the input as a feed dictionary. 
The framework will then automatically handle the model initialization, the construction of the data batches, and the training\&evaluation process.
If extra information is needed for feed-in batches, users can inherit a dataset class from base models to the new model, and define their own training and evaluation batches.
If users plan to introduce a new task, new runners should be defined to configure training and evaluation, and a new reader is needed to accommodate a novel data format if necessary. 
Subsequently, users can develop new models for the novel task or extend existing models to the new task by inheritance and overriding.

In the future, we will continue to update the library to include more models and tasks.

\section{Experiments}
\label{sec:exp}

In this section, we present experiments for three mainly updated tasks in ReChorus2.0: (1)Impression-based ranking and reranking~(Section~\ref{ssec:usageImpression}), (2) CTR Prediction~(Section~\ref{ssec:usageCTR}), and (3) Context-aware Top-k Recommendation~(Section~\ref{ssec:usageTopk}) over all recommendation methods in Section~\ref{ssec:frameworkModel} that can support corresponding tasks.

Through extensive experiments, we aim to demonstrate how ReChorus2.0 flexibly executes various tasks using diverse models on the same datasets, rather than provide a standard benchmark for the recommenders. 

\subsection{Experiment Settings}
\label{ssec:expSetting}

\subsubsection{Datasets}
\label{sssec:expSetData}

\begin{table*}[]
\setlength{\abovecaptionskip}{0cm}  
\setlength{\belowcaptionskip}{-0.2cm} 
\caption{Dataset statistics. \# indicates \textit{the number of }, Inter. is \textit{Interaction}, Imp. is \textit{Impression}, Pos. is \textit{Positive}, Feat. is \textit{Feature}, and Situ. is \textit{Situation}.}
\label{tab:datasetStatistic}
\begin{tabular}{l|rrrrrrrrr}
\toprule
\textbf{Dataset} & \textbf{\#User} & \textbf{\#Item} & \textbf{\#Inter.} & \textbf{\#Positive Inter.} & \textbf{\#Imp.} & \textbf{Imp. Length} & \textbf{\#Pos./Imp.}  & \textbf{\#Item Feat.} & \textbf{\#Situ. Feat.}\\
\midrule
\textbf{MIND} & 269,311  & 9,372 & 67,573,618 & 2,846,791 & 1,689,351 & 39.69 & 1.69 & 3 & 4 \\
\textbf{ML-1M} & 6,034 & 3,125  & 994,292 & 574,286 & 120,562  & 19.679 & 11.552 & 2 & 4 \\
\bottomrule
\end{tabular}
\end{table*}

For consistent comparisons, we choose the same datasets for three task settings, so the datasets should include both rich context information and complete impression logs with exposure and click records.
Based on these requirements, we select two datasets with different scales and scenarios:
\begin{itemize}[nolistsep, leftmargin=*]
    \item \textbf{MIND-Large}~\cite{wu2020mind}: It is a large-scale dataset of impression-based click and exposure logs of Microsoft News from 1 million users during one-week interactions, including rich content and context about news.
    \item \textbf{MovieLens-1M}\footnote{https://grouplens.org/datasets/movielens/1m/}: It contains 1 million rating interactions spanning over 20 years from the MovieLens website with genre and title categories of movie items. 
    
\end{itemize}

Similar preprocesses are conducted for both datasets:
Firstly, five-core filtering on users and items is conducted until all remaining users and items have at least five positive interactions.
Secondly, situation context is extracted from the interaction timestamps, including hour-of-day, day-of-week, and period-of-day.
Thirdly, interactions with the same timestamp are taken as an impression in MIND-Large, and for MovideLens-1M, we follow previous practice~\cite{Jiang2018BeyondGR,Liu2021VariationCA,Ren2023SlateAwareRF} to split user rating sessions into impression lists of length 20 and take the ratings of 4-5 as positive and 1-3 as negative.
Finally, training, validation, and test sets are split along the global timeline~\cite{sun2023take}: In MIND, the first six days are treated as training set, followed by a half-day validation set and half-day test set;
In MovieLens-1M, training, validation, and test sets are split with 80\%, 10\%, 10\% of the time.
The detailed preprocessing scripts are publicly available\footnote{\url{https://github.com/THUwangcy/ReChorus/tree/master/data}},
and final statistics of two datasets are shown in Table~\ref{tab:datasetStatistic}.
We refer to the datasets as \textbf{MIND} and \textbf{ML-1M} for short when the context is clear.

\subsubsection{Training and Evaluation Settings}
\label{sssec:expSetMetric}

Since the three tasks are naturally different, distinct settings are adopted for each task:

\begin{itemize}[nolistsep, leftmargin=*]
\item For impression-based ranking and reranking tasks, list-level BPR loss~\cite{li2023intent} is adopted as training loss, and evaluations are conducted on the ranking/reranking results of impression lists. Evaluation metrics are HR@k, MAP@k, and NDCG@k. Since the impression lists are generally short, results with k=2 and k=5 are reported.
\item For CTR prediction tasks, BCE loss is used for training. 
We evaluate by AUC and Log loss; both are commonly used evaluation metrics for CTR prediction.
\item For context-aware Top-k recommendation, negative sampling is utilized for both training and evaluation: one random negative sample is chosen for BPR loss during training, and 99 negative samples are randomly sampled for evaluation, with HR@k and NDCG@k as metrics~(k=5, 10, and 20 are reported).
\end{itemize}

Note that for Top-k recommendations, efficiency considerations led us to sample only 99 negative instances due to the necessity of running and tuning numerous models. However, ReChorus2.0 supports evaluation with more negative instances or even the full item set.


\begin{table}[]
\setlength{\abovecaptionskip}{0cm}  
\setlength{\belowcaptionskip}{-0.2cm} 
\caption{Performances of ranking and re-ranking models on impressions of MIND dataset. N. is short for \textit{NDCG}, and SetR. is short for the \textit{SetRank} model.}
\label{tab:resultRankMind}
\small
\resizebox{1.01\linewidth}{!}{
\begin{tabular}{cc|cccccc}
\toprule
& \textbf{Model} & HR@2 & MAP@2 & N.@2 & HR@5 & MAP@5 & N.@5 \\
 \midrule
\parbox[t]{1mm}{\multirow{3}{*}{\rotatebox[origin=c]{90}{\textbf{Ranking}}}} & BPR & 0.3887 & 0.2598 & 0.2888 & 0.6476 & 0.3191 & 0.3953 \\
& GRU4Rec & 0.4406 & 0.3049 & 0.3356 & 0.6752 & 0.3547 & 0.4297 \\
& SASRec & 0.4341 & 0.3005 & 0.3307 & 0.6683 & 0.3495 & 0.4240 \\
\midrule
\parbox[t]{1mm}{\multirow{6}{*}{\rotatebox[origin=c]{90}{\textbf{ReRanking}}}} & PRM+BPR & 0.4443 & 0.3249 & 0.3522 & 0.6654 & 0.3660 & 0.4363 \\
& SetR.+BPR & 0.4445  & 0.3241 & 0.3516 & 0.6663 & 0.3653 & 0.4359 \\
& MIR+BPR & 0.4661 & 0.3377 & 0.3670 & 0.6920 & 0.3845 & 0.4578 \\
& PRM+GRU & 0.4558 & 0.3317 & 0.3601 & 0.6697 & 0.3714 & 0.4416 \\
& SetR.+GRU & 0.4441 & 0.3215 & 0.3494 & 0.6656 & 0.3629 & 0.4337 \\
& MIR+GRU & 0.4722  & 0.3432  & 0.3727 & 0.6997 & 0.3915 & 0.4653 \\
\bottomrule
\end{tabular}
}
\end{table}

\begin{table}[]
\setlength{\abovecaptionskip}{0cm}  
\setlength{\belowcaptionskip}{-0.2cm} 
\caption{Performances of ranking and re-ranking models on impressions of ML-1M dataset. GCN and SAS represent LightGCN and SASRec as backbone. All other notations are the same as Table~\ref{tab:resultRankMind}.}
\label{tab:resultRankML}
\small
\resizebox{1.01\linewidth}{!}{
\begin{tabular}{cc|cccccc}
\toprule
& \textbf{Model} & HR@2 & MAP@2 & N.@2 & HR@5 & MAP@5 & N.@5 \\
 \midrule
\parbox[t]{1mm}{\multirow{4}{*}{\rotatebox[origin=c]{90}{\textbf{Ranking}}}} & BPR & 0.8930 & 0.6943 & 0.7347 & 0.9734 & 0.6605 & 0.7449 \\
& LightGCN &  0.8817 & 0.6978 & 0.7358 & 0.9835  & 0.6690  & 0.7538 \\
& GRU4Rec & 0.8822 & 0.6859 & 0.7257 & 0.9816 & 0.6333 & 0.7240 \\
& SASRec & 0.8920 & 0.7046 & 0.7431 & 0.9767  & 0.6556  & 0.7431 \\
\midrule
\parbox[t]{1mm}{\multirow{6}{*}{\rotatebox[origin=c]{90}{\textbf{ReRanking}}}} & PRM+GCN &  0.8816  & 0.7073 & 0.7436 & 0.9842 & 0.6606  & 0.7475 \\
& SetR.+GCN & 0.8853 & 0.7057  & 0.7410  & 0.9842 & 0.6626 & 0.7490 \\
& MIR+GCN & 0.8724 & 0.7022 & 0.7374 & 0.9926 & 0.6544  & 0.7433 \\
& PRM+SAS & 0.8971 & 0.7095 & 0.7459 & 0.9890 & 0.6539 & 0.7441 \\
& SetR.+SAS & 0.8908 & 0.7098 & 0.7467 & 0.9877 & 0.6595 & 0.7479 \\
& MIR+SAS & 0.8920 & 0.6994 & 0.7485 & 0.9853 & 0.6474 & 0.7379 \\
\bottomrule
\end{tabular}
}
\end{table}

\subsubsection{Implementation Details}
\label{sssec:expSetImplement}

ReChorus2.0 is implemented entirely in Python and based on PyTorch. For a lightweight interface, all user interactions are facilitated via command lines. 
Parameter tuning is conducted for each model and task: for fair comparison, the embedding size is fixed at 64; learning rate is chosen from \{5e-3, 2e-3, 1e-3, 5e-4, 2e-4\}; L2 regularization term is chosen from \{1e-4, 1e-6, 0\}; for models with hidden layers, the size of each hidden layer is chosen from \{64, 128, 256\}; the length of user history for sequential models is selected from \{10, 20, 30\}; and other model-specific parameters are also fine-tuned accordingly. 
Training employs early stopping, which stops when there is no performance improvement on the validation set for 10 epochs.

The optimal parameters for each model on each task are recorded within the library\footnote{\url{https://github.com/THUwangcy/ReChorus/tree/master/docs/demo_scripts_results}}. Experiments were repeated five times with different random seeds using the \textit{exp.py} module, and the average performance is reported as follows.

\subsection{Experiment Results}
\label{ssec:expResult}
\begin{table*}[]
\setlength{\abovecaptionskip}{0cm}  
\setlength{\belowcaptionskip}{-0.2cm} 
\caption{Context-aware recommender systems~(CARS) for the top-k recommendation task.}
\label{tab:resultCARStopk}
\small
\begin{tabular}{c|rrrrrr|rrrrrr}
\toprule
\multirow{2}{*}{\textbf{\textbf{Model}}} & \multicolumn{6}{c|}{\textbf{MIND}} & \multicolumn{6}{c}{\textbf{ML-1M}} \\
\textbf{} & HR@5 & NDCG@5 & HR@10 & NDCG@10 & HR@20 & NDCG@20 & HR@5 & NDCG@5 & HR@10 & NDCG@10 & HR@20 & NDCG@20 \\
\midrule
FM & 0.1204 & 0.0745 & 0.1923 & 0.0977 & 0.2663 & 0.1164 & 0.3214 & 0.2113 & 0.4870 & 0.2646 & 0.6836 & 0.3142 \\
Wide\&Deep & 0.0949 & 0.0623 & 0.1339 & 0.0749 & 0.1809 & 0.0868 & 0.3641 & 0.2462 & 0.5289 & 0.2991 & 0.7248 & 0.3486 \\
DeepFM & 0.1183 & 0.0780 & 0.1942 & 0.1022 & 0.3018 & 0.1295  & 0.3264 & 0.2180 & 0.4866 & 0.2694 & 0.6853 & 0.3195 \\
AFM & 0.1199 & 0.0762 & 0.1650 & 0.0908 & 0.2325 & 0.1077 & 0.3845 & 0.2600 & 0.5538 & 0.3146 & 0.7551 & 0.3656 \\
DCN & 0.1146 & 0.0771 & 0.1725 & 0.0957 & 0.2660 & 0.1191 & 0.3695 & 0.2503 & 0.5370 & 0.3042 & 0.7288 & 0.3527 \\
xDeepFM & 0.1143 & 0.0712 & 0.1787 & 0.0919 & 0.2792 & 0.1171 & 0.3320 & 0.2234 & 0.4930 & 0.2751 & 0.6847 & 0.3235 \\
AutoInt & 0.1384 & 0.0927 & 0.2151 & 0.1172 & 0.3594 & 0.1533 & 0.3615 & 0.2439 & 0.5205 & 0.2951 & 0.7150 & 0.3443 \\
DCNv2 & 0.1472 & 0.0978 & 0.2222 & 0.1218 & 0.3666 & 0.1577 & 0.3740 & 0.2541 & 0.5473 & 0.3100 & 0.7403 & 0.3588 \\
FinalMLP & 0.1759 & 0.1161 & 0.2558 & 0.1419 & 0.3576 & 0.1676 & 0.3834 & 0.2627 & 0.5464 & 0.3152 & 0.7404 & 0.3643 \\
SAM & 0.0935 & 0.0579 & 0.1566 & 0.0782 & 0.2442 & 0.1002 & 0.3774 & 0.2564 & 0.5451 & 0.3104 & 0.7434 & 0.3605 \\
\midrule
DIN & 0.2536 & 0.1686  & 0.3627 & 0.2038 & 0.4030 & 0.2390 & 0.4939 & 0.3615 & 0.6448 & 0.4101 & 0.7921 & 0.4475 \\
DIEN & 0.2351 & 0.1628 & 0.2946 & 0.1822 & 0.3543 & 0.1970  & 0.5227 & 0.3881 & 0.6705 & 0.4360 & 0.8163 & 0.4729 \\
CAN & 0.1952 & 0.1203 & 0.3041 & 0.1622 & 0.4946 & 0.1999 & 0.4989 & 0.3624 & 0.6441 & 0.4110 & 0.7958 & 0.4517 \\
\bottomrule
\end{tabular}
\end{table*}

\begin{table}[]
\setlength{\abovecaptionskip}{0cm}  
\setlength{\belowcaptionskip}{-0.2cm} 
\caption{Context-aware recommender systems~(CARS) for the CTR prediction task.}
\label{tab:resultCARS}
\begin{tabular}{c|rr|rr}
\toprule
\textbf{\textbf{Model}} & \multicolumn{2}{c|}{MIND} & \multicolumn{2}{c}{ML-1M} \\
\textbf{} & AUC ($\uparrow$) & Log loss ($\downarrow$) & AUC ($\uparrow$) & Log loss ($\downarrow$) \\
\midrule
FM & 0.6386 & 0.1661 & 0.7661 & 0.9155 \\
Wide\&Deep & 0.6525 & 0.1582 & 0.7798 & 0.6291 \\
DeepFM & 0.6493 & 0.1678 & 0.7736 & 0.6252 \\
AFM & 0.6476 & 0.1572 & 0.7795 & 0.7564 \\
DCN & 0.6513 & 0.1587 & 0.7831 & 0.6334 \\
xDeepFM & 0.6494 & 0.1593 & 0.7692 &  0.6203 \\
AutoInt & 0.6547 & 0.1578 & 0.7785 & 0.5661 \\
DCNv2 & 0.6518 & 0.1642 & 0.7870 & 0.5572 \\
FinalMLP & 0.6484 & 0.1614 & 0.7841 & 0.5735 \\
SAM & 0.6508 & 0.1707 & 0.7919 & 0.5590 \\
\midrule
DIN & 0.6625 & 0.1609 & 0.7904 & 0.6052 \\
DIEN & 0.6604 & 0.1656  & 0.7874 & 0.5513 \\
CAN & 0.6598 & 0.1709 & 0.7909 & 0.5510 \\
\bottomrule
\end{tabular}
\end{table}

\subsubsection{Impression-based Ranking\&Re-ranking}
Table~\ref{tab:resultRankMind} and Table~\ref{tab:resultRankML} shows the performance comparison for ranking and re-ranking models on the MIND and ML-1M dataset, respectively. 
For ranking models, we consider two classical general models, BPR and LightGCN, and two sequential models, GRU4Rec and SASRec.
For the reranking task, each reranker is conducted with one general model and one sequential model as base rankers, respectively, where base rankers are all previously trained on the same dataset and fixed.

In the MIND dataset, reranking the candidate item list improves the ranking performance in most cases, especially for BPRMF, the simpler base ranker.
Sequential base rankers can also provide more useful information, and they tend to achieve continuously better performance for every reranking model.
The promising results indicate that considering the candidate list context induces a performance boost.
For the ML-1M dataset, the candidate list is shorter, and the proportion of positive interactions is much larger on average, which makes the ranking task easier~(i.e., the metrics are higher for base rankers). 
Hence, limited improvement can be achieved by adding sequential information and reranking models.

\subsubsection{CARS for Top-k Recommendation and CTR Prediction}
For the Context-aware recommender systems~(CARS) mentioned in Section~\ref{ssec:frameworkModel}, we investigate their performance on both Top-k recommendation and CTR prediction tasks in Table~\ref{tab:resultCARStopk} and Table~\ref{tab:resultCARS}. 
Models are divided into two group by whether they utilize user history, and they are listed in chronological order in each group.

Despite the difference in the data size and the ratio of positive interactions between the two datasets, Table~\ref{tab:resultCARStopk} reveals that the sequential CARS achieves much better performance in Top-k recommendation. Apart from the information gained from the historical interaction, this result is also explained by the fact that sequential models can better leverage the newly emerged interactions in the validation and test set. 
The results of CARS for the CTR prediction task are shown in Table~\ref{tab:resultCARS}, where sequential recommenders also achieve slightly better performance on average. 
Note that the \textit{Log loss} metric is not compatible between two datasets: Because the ratio of positive interactions is much lower for the MIND dataset, models are prone to give smaller prediction scores to achieve smaller log loss values.

\vspace{0.2cm}
The above experiment results for various recommenders on the three newly implemented tasks highlight the flexibility of ReChorus2.0 framework, demonstrating its ability to handle various types of tasks, inputs, settings, and metrics comprehensively.

\section{Conclusions}
\label{sec:conclusion}

In this paper, we provide \textbf{ReChorus2.0}, a modular task-flexible recommender library.
Different from existing popular libraries, 
ReChorus2.0 provides highly customized ways to implement tasks and consturct datasets.
ReChorus2.0 concurrently supports three types of recommendation tasks: top-k recommendation, CTR prediction, and impression-based ranking/re-ranking. It also enables the same recommender to flexibly support multiple different tasks. 
Moreover, utilizing the well-separated reader module, ReChorus2.0 accommodates customized input settings, including various dataset splitting configurations,
different interaction formats~(e.g., negative item lists, impression logs, and click labels), 
and various types of context information.
With these characteristics, ReChorus2.0 offers researchers in the recommender system scenario a convenient platform that enables them to assemble different modules like building blocks, facilitating the creation of customized experiment settings. 
This is particularly crucial for users to conduct in-depth analyses of methods or data during the research process.
In the future, we plan to provide more task configurations and models, offering users a more convenient, powerful, and comprehensive tool for recommender system research.

\begin{acks}
We sincerely thank Chenyang Wang's excellent contributions to ReChorus1.0.
It is his elegant work that makes ReChorus2.0 possible.
This work is supported by the Natural Science Foundation of China (Grant No. U21B2026, 62372260) and Quan Cheng Laboratory (Grant No. QCLZD202301).
\end{acks}

\clearpage

\bibliographystyle{ACM-Reference-Format}
\bibliography{reference}


\begin{thebibliography}{71}


\ifx \showCODEN    \undefined \def \showCODEN     #1{\unskip}     \fi
\ifx \showDOI      \undefined \def \showDOI       #1{#1}\fi
\ifx \showISBNx    \undefined \def \showISBNx     #1{\unskip}     \fi
\ifx \showISBNxiii \undefined \def \showISBNxiii  #1{\unskip}     \fi
\ifx \showISSN     \undefined \def \showISSN      #1{\unskip}     \fi
\ifx \showLCCN     \undefined \def \showLCCN      #1{\unskip}     \fi
\ifx \shownote     \undefined \def \shownote      #1{#1}          \fi
\ifx \showarticletitle \undefined \def \showarticletitle #1{#1}   \fi
\ifx \showURL      \undefined \def \showURL       {\relax}        \fi
\providecommand\bibfield[2]{#2}
\providecommand\bibinfo[2]{#2}
\providecommand\natexlab[1]{#1}
\providecommand\showeprint[2][]{arXiv:#2}

\bibitem[Ai et~al\mbox{.}(2018)]%
        {ai_learning_2018}
\bibfield{author}{\bibinfo{person}{Qingyao Ai}, \bibinfo{person}{Keping Bi}, \bibinfo{person}{Jiafeng Guo}, {and} \bibinfo{person}{W.~Bruce Croft}.} \bibinfo{year}{2018}\natexlab{}.
\newblock \showarticletitle{Learning a {Deep} {Listwise} {Context} {Model} for {Ranking} {Refinement}}. In \bibinfo{booktitle}{\emph{SIGIR}}. \bibinfo{publisher}{ACM}, \bibinfo{address}{Ann Arbor MI USA}, \bibinfo{pages}{135--144}.
\newblock


\bibitem[Anelli et~al\mbox{.}(2021)]%
        {Vito2021elliot}
\bibfield{author}{\bibinfo{person}{Vito~Walter Anelli}, \bibinfo{person}{Alejandro Bellog{\'{\i}}n}, \bibinfo{person}{Antonio Ferrara}, \bibinfo{person}{Daniele Malitesta}, \bibinfo{person}{Felice~Antonio Merra}, \bibinfo{person}{Claudio Pomo}, \bibinfo{person}{Francesco~Maria Donini}, {and} \bibinfo{person}{Tommaso~Di Noia}.} \bibinfo{year}{2021}\natexlab{}.
\newblock \showarticletitle{Elliot: {A} Comprehensive and Rigorous Framework for Reproducible Recommender Systems Evaluation}. In \bibinfo{booktitle}{\emph{{SIGIR} '21: The 44th International {ACM} {SIGIR} Conference on Research and Development in Information Retrieval, Virtual Event, Canada, July 11-15, 2021}}, \bibfield{editor}{\bibinfo{person}{Fernando Diaz}, \bibinfo{person}{Chirag Shah}, \bibinfo{person}{Torsten Suel}, \bibinfo{person}{Pablo Castells}, \bibinfo{person}{Rosie Jones}, {and} \bibinfo{person}{Tetsuya Sakai}} (Eds.). \bibinfo{publisher}{{ACM}}, \bibinfo{pages}{2405--2414}.
\newblock
\urldef\tempurl%
\url{https://doi.org/10.1145/3404835.3463245}
\showDOI{\tempurl}


\bibitem[Argyriou et~al\mbox{.}(2020)]%
        {argyriou2020microsoft}
\bibfield{author}{\bibinfo{person}{Andreas Argyriou}, \bibinfo{person}{Miguel Gonz{\'a}lez-Fierro}, {and} \bibinfo{person}{Le Zhang}.} \bibinfo{year}{2020}\natexlab{}.
\newblock \showarticletitle{Microsoft recommenders: best practices for production-ready recommendation systems}. In \bibinfo{booktitle}{\emph{Companion Proceedings of the Web Conference 2020}}. \bibinfo{pages}{50--51}.
\newblock


\bibitem[Bian et~al\mbox{.}(2022)]%
        {bian2022can}
\bibfield{author}{\bibinfo{person}{Weijie Bian}, \bibinfo{person}{Kailun Wu}, \bibinfo{person}{Lejian Ren}, \bibinfo{person}{Qi Pi}, \bibinfo{person}{Yujing Zhang}, \bibinfo{person}{Can Xiao}, \bibinfo{person}{Xiang-Rong Sheng}, \bibinfo{person}{Yong-Nan Zhu}, \bibinfo{person}{Zhangming Chan}, \bibinfo{person}{Na Mou}, {et~al\mbox{.}}} \bibinfo{year}{2022}\natexlab{}.
\newblock \showarticletitle{CAN: feature co-action network for click-through rate prediction}. In \bibinfo{booktitle}{\emph{Proceedings of the fifteenth ACM international conference on web search and data mining}}. \bibinfo{pages}{57--65}.
\newblock


\bibitem[Cao et~al\mbox{.}(2022)]%
        {cao2022sdim}
\bibfield{author}{\bibinfo{person}{Yue Cao}, \bibinfo{person}{Xiaojiang Zhou}, \bibinfo{person}{Jiaqi Feng}, \bibinfo{person}{Peihao Huang}, \bibinfo{person}{Yao Xiao}, \bibinfo{person}{Dayao Chen}, {and} \bibinfo{person}{Sheng Chen}.} \bibinfo{year}{2022}\natexlab{}.
\newblock \showarticletitle{Sampling is all you need on modeling long-term user behaviors for CTR prediction}. In \bibinfo{booktitle}{\emph{Proceedings of the 31st ACM International Conference on Information \& Knowledge Management}}. \bibinfo{pages}{2974--2983}.
\newblock


\bibitem[Cao et~al\mbox{.}(2007)]%
        {Cao2007LearningTR}
\bibfield{author}{\bibinfo{person}{Zhe Cao}, \bibinfo{person}{Tao Qin}, \bibinfo{person}{Tie-Yan Liu}, \bibinfo{person}{Ming-Feng Tsai}, {and} \bibinfo{person}{Hang Li}.} \bibinfo{year}{2007}\natexlab{}.
\newblock \showarticletitle{Learning to rank: from pairwise approach to listwise approach}. In \bibinfo{booktitle}{\emph{International Conference on Machine Learning}}.
\newblock
\urldef\tempurl%
\url{https://api.semanticscholar.org/CorpusID:207163577}
\showURL{%
\tempurl}


\bibitem[Cen et~al\mbox{.}(2020)]%
        {cen2020comirec}
\bibfield{author}{\bibinfo{person}{Yukuo Cen}, \bibinfo{person}{Jianwei Zhang}, \bibinfo{person}{Xu Zou}, \bibinfo{person}{Chang Zhou}, \bibinfo{person}{Hongxia Yang}, {and} \bibinfo{person}{Jie Tang}.} \bibinfo{year}{2020}\natexlab{}.
\newblock \showarticletitle{Controllable multi-interest framework for recommendation}. In \bibinfo{booktitle}{\emph{Proceedings of the 26th ACM SIGKDD International Conference on Knowledge Discovery \& Data Mining}}. \bibinfo{pages}{2942--2951}.
\newblock


\bibitem[Chen et~al\mbox{.}(2021)]%
        {chen2021eta}
\bibfield{author}{\bibinfo{person}{Qiwei Chen}, \bibinfo{person}{Changhua Pei}, \bibinfo{person}{Shanshan Lv}, \bibinfo{person}{Chao Li}, \bibinfo{person}{Junfeng Ge}, {and} \bibinfo{person}{Wenwu Ou}.} \bibinfo{year}{2021}\natexlab{}.
\newblock \showarticletitle{End-to-end user behavior retrieval in click-through rateprediction model}.
\newblock \bibinfo{journal}{\emph{arXiv preprint arXiv:2108.04468}} (\bibinfo{year}{2021}).
\newblock


\bibitem[Cheng et~al\mbox{.}(2016)]%
        {cheng2016widedeep}
\bibfield{author}{\bibinfo{person}{Heng-Tze Cheng}, \bibinfo{person}{Levent Koc}, \bibinfo{person}{Jeremiah Harmsen}, \bibinfo{person}{Tal Shaked}, \bibinfo{person}{Tushar Chandra}, \bibinfo{person}{Hrishi Aradhye}, \bibinfo{person}{Glen Anderson}, \bibinfo{person}{Greg Corrado}, \bibinfo{person}{Wei Chai}, \bibinfo{person}{Mustafa Ispir}, {et~al\mbox{.}}} \bibinfo{year}{2016}\natexlab{}.
\newblock \showarticletitle{Wide \& deep learning for recommender systems}. In \bibinfo{booktitle}{\emph{Proceedings of the 1st workshop on deep learning for recommender systems}}. \bibinfo{pages}{7--10}.
\newblock


\bibitem[Cheng and Xue(2021)]%
        {cheng2021sam}
\bibfield{author}{\bibinfo{person}{Yuan Cheng} {and} \bibinfo{person}{Yanbo Xue}.} \bibinfo{year}{2021}\natexlab{}.
\newblock \showarticletitle{Looking at CTR Prediction Again: Is Attention All You Need?}. In \bibinfo{booktitle}{\emph{Proceedings of the 44th International ACM SIGIR Conference on Research and Development in Information Retrieval}}. \bibinfo{pages}{1279--1287}.
\newblock


\bibitem[Chia et~al\mbox{.}(2022)]%
        {chia2022RecList}
\bibfield{author}{\bibinfo{person}{Patrick~John Chia}, \bibinfo{person}{Jacopo Tagliabue}, \bibinfo{person}{Federico Bianchi}, \bibinfo{person}{Chloe He}, {and} \bibinfo{person}{Brian Ko}.} \bibinfo{year}{2022}\natexlab{}.
\newblock \showarticletitle{Beyond NDCG: Behavioral Testing of Recommender Systems with RecList} \emph{(\bibinfo{series}{WWW '22 Companion})}. \bibinfo{publisher}{Association for Computing Machinery}, \bibinfo{address}{New York, NY, USA}, \bibinfo{pages}{99–104}.
\newblock
\showISBNx{9781450391306}
\urldef\tempurl%
\url{https://doi.org/10.1145/3487553.3524215}
\showDOI{\tempurl}


\bibitem[Ekstrand(2020)]%
        {ekstrand2020lenskit}
\bibfield{author}{\bibinfo{person}{Michael~D Ekstrand}.} \bibinfo{year}{2020}\natexlab{}.
\newblock \showarticletitle{Lenskit for python: Next-generation software for recommender systems experiments}. In \bibinfo{booktitle}{\emph{Proceedings of the 29th ACM international conference on information \& knowledge management}}. \bibinfo{pages}{2999--3006}.
\newblock


\bibitem[Feng et~al\mbox{.}(2021)]%
        {feng_grn_2021}
\bibfield{author}{\bibinfo{person}{Yufei Feng}, \bibinfo{person}{Binbin Hu}, \bibinfo{person}{Yu Gong}, \bibinfo{person}{Fei Sun}, \bibinfo{person}{Qingwen Liu}, {and} \bibinfo{person}{Wenwu Ou}.} \bibinfo{year}{2021}\natexlab{}.
\newblock \showarticletitle{{GRN}: {Generative} {Rerank} {Network} for {Context}-wise {Recommendation}}.
\newblock \bibinfo{journal}{\emph{arXiv}} (\bibinfo{date}{April} \bibinfo{year}{2021}).
\newblock
\newblock
\shownote{arXiv}.


\bibitem[Ferrari~Dacrema et~al\mbox{.}(2021)]%
        {ferrari2021troubling}
\bibfield{author}{\bibinfo{person}{Maurizio Ferrari~Dacrema}, \bibinfo{person}{Simone Boglio}, \bibinfo{person}{Paolo Cremonesi}, {and} \bibinfo{person}{Dietmar Jannach}.} \bibinfo{year}{2021}\natexlab{}.
\newblock \showarticletitle{A troubling analysis of reproducibility and progress in recommender systems research}.
\newblock \bibinfo{journal}{\emph{ACM Transactions on Information Systems (TOIS)}} \bibinfo{volume}{39}, \bibinfo{number}{2} (\bibinfo{year}{2021}), \bibinfo{pages}{1--49}.
\newblock


\bibitem[Guo et~al\mbox{.}(2017)]%
        {guo2017deepfm}
\bibfield{author}{\bibinfo{person}{Huifeng Guo}, \bibinfo{person}{Ruiming Tang}, \bibinfo{person}{Yunming Ye}, \bibinfo{person}{Zhenguo Li}, {and} \bibinfo{person}{Xiuqiang He}.} \bibinfo{year}{2017}\natexlab{}.
\newblock \showarticletitle{DeepFM: a factorization-machine based neural network for CTR prediction}.
\newblock \bibinfo{journal}{\emph{arXiv preprint arXiv:1703.04247}} (\bibinfo{year}{2017}).
\newblock


\bibitem[He et~al\mbox{.}(2020)]%
        {he2020lightgcn}
\bibfield{author}{\bibinfo{person}{Xiangnan He}, \bibinfo{person}{Kuan Deng}, \bibinfo{person}{Xiang Wang}, \bibinfo{person}{Yan Li}, \bibinfo{person}{Yongdong Zhang}, {and} \bibinfo{person}{Meng Wang}.} \bibinfo{year}{2020}\natexlab{}.
\newblock \showarticletitle{Lightgcn: Simplifying and powering graph convolution network for recommendation}. In \bibinfo{booktitle}{\emph{Proceedings of the 43rd International ACM SIGIR conference on research and development in Information Retrieval}}. \bibinfo{pages}{639--648}.
\newblock


\bibitem[He et~al\mbox{.}(2017)]%
        {he2017neuCF}
\bibfield{author}{\bibinfo{person}{Xiangnan He}, \bibinfo{person}{Lizi Liao}, \bibinfo{person}{Hanwang Zhang}, \bibinfo{person}{Liqiang Nie}, \bibinfo{person}{Xia Hu}, {and} \bibinfo{person}{Tat-Seng Chua}.} \bibinfo{year}{2017}\natexlab{}.
\newblock \showarticletitle{Neural collaborative filtering}. In \bibinfo{booktitle}{\emph{Proceedings of the 26th international conference on world wide web}}. \bibinfo{pages}{173--182}.
\newblock


\bibitem[Hidasi et~al\mbox{.}(2015)]%
        {hidasi2015gru4rec}
\bibfield{author}{\bibinfo{person}{Bal{\'a}zs Hidasi}, \bibinfo{person}{Alexandros Karatzoglou}, \bibinfo{person}{Linas Baltrunas}, {and} \bibinfo{person}{Domonkos Tikk}.} \bibinfo{year}{2015}\natexlab{}.
\newblock \showarticletitle{Session-based recommendations with recurrent neural networks}.
\newblock \bibinfo{journal}{\emph{arXiv preprint arXiv:1511.06939}} (\bibinfo{year}{2015}).
\newblock


\bibitem[Huzhang et~al\mbox{.}(2020)]%
        {huzhang_aliexpress_2020}
\bibfield{author}{\bibinfo{person}{Guangda Huzhang}, \bibinfo{person}{Zhen-Jia Pang}, \bibinfo{person}{Yongqing Gao}, \bibinfo{person}{Yawen Liu}, \bibinfo{person}{Weijie Shen}, \bibinfo{person}{Wen-Ji Zhou}, \bibinfo{person}{Qing Da}, \bibinfo{person}{An-Xiang Zeng}, \bibinfo{person}{Han Yu}, \bibinfo{person}{Yang Yu}, {and} \bibinfo{person}{Zhi-Hua Zhou}.} \bibinfo{year}{2020}\natexlab{}.
\newblock \showarticletitle{{AliExpress} {Learning}-{To}-{Rank}: {Maximizing} {Online} {Model} {Performance} without {Going} {Online}}.
\newblock \bibinfo{journal}{\emph{arXiv}} (\bibinfo{date}{Dec.} \bibinfo{year}{2020}).
\newblock


\bibitem[Iana et~al\mbox{.}(2023)]%
        {iana2023newsreclib}
\bibfield{author}{\bibinfo{person}{Andreea Iana}, \bibinfo{person}{Goran Glava{\v{s}}}, {and} \bibinfo{person}{Heiko Paulheim}.} \bibinfo{year}{2023}\natexlab{}.
\newblock \showarticletitle{NewsRecLib: A PyTorch-Lightning Library for Neural News Recommendation}. In \bibinfo{booktitle}{\emph{Proceedings of the 2023 Conference on Empirical Methods in Natural Language Processing: System Demonstrations}}. \bibinfo{pages}{296--310}.
\newblock


\bibitem[J{\"a}rvelin and Kek{\"a}l{\"a}inen(2002)]%
        {jarvelin2002cumulated}
\bibfield{author}{\bibinfo{person}{Kalervo J{\"a}rvelin} {and} \bibinfo{person}{Jaana Kek{\"a}l{\"a}inen}.} \bibinfo{year}{2002}\natexlab{}.
\newblock \showarticletitle{Cumulated gain-based evaluation of IR techniques}.
\newblock \bibinfo{journal}{\emph{ACM Transactions on Information Systems (TOIS)}} \bibinfo{volume}{20}, \bibinfo{number}{4} (\bibinfo{year}{2002}), \bibinfo{pages}{422--446}.
\newblock


\bibitem[Jiang et~al\mbox{.}(2018)]%
        {Jiang2018BeyondGR}
\bibfield{author}{\bibinfo{person}{Ray Jiang}, \bibinfo{person}{Sven Gowal}, \bibinfo{person}{Yuqiu Qian}, \bibinfo{person}{Timothy~A. Mann}, {and} \bibinfo{person}{Danilo~Jimenez Rezende}.} \bibinfo{year}{2018}\natexlab{}.
\newblock \showarticletitle{Beyond Greedy Ranking: Slate Optimization via List-CVAE}. In \bibinfo{booktitle}{\emph{International Conference on Learning Representations}}.
\newblock
\urldef\tempurl%
\url{https://api.semanticscholar.org/CorpusID:44005913}
\showURL{%
\tempurl}


\bibitem[Kang and McAuley(2018)]%
        {kang2018sasrec}
\bibfield{author}{\bibinfo{person}{Wang-Cheng Kang} {and} \bibinfo{person}{Julian McAuley}.} \bibinfo{year}{2018}\natexlab{}.
\newblock \showarticletitle{Self-attentive sequential recommendation}. In \bibinfo{booktitle}{\emph{2018 IEEE international conference on data mining (ICDM)}}. IEEE, \bibinfo{pages}{197--206}.
\newblock


\bibitem[Koren et~al\mbox{.}(2009)]%
        {koren2009matrix}
\bibfield{author}{\bibinfo{person}{Yehuda Koren}, \bibinfo{person}{Robert Bell}, {and} \bibinfo{person}{Chris Volinsky}.} \bibinfo{year}{2009}\natexlab{}.
\newblock \showarticletitle{Matrix factorization techniques for recommender systems}.
\newblock \bibinfo{journal}{\emph{Computer}} \bibinfo{volume}{42}, \bibinfo{number}{8} (\bibinfo{year}{2009}), \bibinfo{pages}{30--37}.
\newblock


\bibitem[Kulkarni and Rodd(2020)]%
        {kulkarni2020context}
\bibfield{author}{\bibinfo{person}{Saurabh Kulkarni} {and} \bibinfo{person}{Sunil~F Rodd}.} \bibinfo{year}{2020}\natexlab{}.
\newblock \showarticletitle{Context Aware Recommendation Systems: A review of the state of the art techniques}.
\newblock \bibinfo{journal}{\emph{Computer Science Review}}  \bibinfo{volume}{37} (\bibinfo{year}{2020}), \bibinfo{pages}{100255}.
\newblock


\bibitem[Lee et~al\mbox{.}(2021)]%
        {lee2021buir}
\bibfield{author}{\bibinfo{person}{Dongha Lee}, \bibinfo{person}{SeongKu Kang}, \bibinfo{person}{Hyunjun Ju}, \bibinfo{person}{Chanyoung Park}, {and} \bibinfo{person}{Hwanjo Yu}.} \bibinfo{year}{2021}\natexlab{}.
\newblock \showarticletitle{Bootstrapping user and item representations for one-class collaborative filtering}. In \bibinfo{booktitle}{\emph{Proceedings of the 44th international ACM SIGIR conference on Research and Development in information retrieval}}. \bibinfo{pages}{317--326}.
\newblock


\bibitem[Li et~al\mbox{.}(2017)]%
        {li2017narm}
\bibfield{author}{\bibinfo{person}{Jing Li}, \bibinfo{person}{Pengjie Ren}, \bibinfo{person}{Zhumin Chen}, \bibinfo{person}{Zhaochun Ren}, \bibinfo{person}{Tao Lian}, {and} \bibinfo{person}{Jun Ma}.} \bibinfo{year}{2017}\natexlab{}.
\newblock \showarticletitle{Neural attentive session-based recommendation}. In \bibinfo{booktitle}{\emph{Proceedings of the 2017 ACM on Conference on Information and Knowledge Management}}. \bibinfo{pages}{1419--1428}.
\newblock


\bibitem[Li et~al\mbox{.}(2024)]%
        {li2024sare}
\bibfield{author}{\bibinfo{person}{Jiayu Li}, \bibinfo{person}{Peijie Sun}, \bibinfo{person}{Chumeng Jiang}, \bibinfo{person}{Weizhi Ma}, \bibinfo{person}{Qingyao Ai}, {and} \bibinfo{person}{Min Zhang}.} \bibinfo{year}{2024}\natexlab{}.
\newblock \showarticletitle{A Situation-aware Enhancer for Personalized Recommendation}.
\newblock \bibinfo{journal}{\emph{arXiv preprint arXiv:2403.18317}} (\bibinfo{year}{2024}).
\newblock


\bibitem[Li et~al\mbox{.}(2023)]%
        {li2023intent}
\bibfield{author}{\bibinfo{person}{Jiayu Li}, \bibinfo{person}{Peijie Sun}, \bibinfo{person}{Zhefan Wang}, \bibinfo{person}{Weizhi Ma}, \bibinfo{person}{Yangkun Li}, \bibinfo{person}{Min Zhang}, \bibinfo{person}{Zhoutian Feng}, {and} \bibinfo{person}{Daiyue Xue}.} \bibinfo{year}{2023}\natexlab{}.
\newblock \showarticletitle{Intent-aware Ranking Ensemble for Personalized Recommendation}. In \bibinfo{booktitle}{\emph{Proceedings of the 46th International ACM SIGIR Conference on Research and Development in Information Retrieval}}. \bibinfo{pages}{1004--1013}.
\newblock


\bibitem[Li et~al\mbox{.}(2020)]%
        {li2020tisasrec}
\bibfield{author}{\bibinfo{person}{Jiacheng Li}, \bibinfo{person}{Yujie Wang}, {and} \bibinfo{person}{Julian McAuley}.} \bibinfo{year}{2020}\natexlab{}.
\newblock \showarticletitle{Time interval aware self-attention for sequential recommendation}. In \bibinfo{booktitle}{\emph{Proceedings of the 13th international conference on web search and data mining}}. \bibinfo{pages}{322--330}.
\newblock


\bibitem[Lian et~al\mbox{.}(2018)]%
        {lian2018xdeepfm}
\bibfield{author}{\bibinfo{person}{Jianxun Lian}, \bibinfo{person}{Xiaohuan Zhou}, \bibinfo{person}{Fuzheng Zhang}, \bibinfo{person}{Zhongxia Chen}, \bibinfo{person}{Xing Xie}, {and} \bibinfo{person}{Guangzhong Sun}.} \bibinfo{year}{2018}\natexlab{}.
\newblock \showarticletitle{xdeepfm: Combining explicit and implicit feature interactions for recommender systems}. In \bibinfo{booktitle}{\emph{Proceedings of the 24th ACM SIGKDD international conference on knowledge discovery \& data mining}}. \bibinfo{pages}{1754--1763}.
\newblock


\bibitem[Liu et~al\mbox{.}(2021)]%
        {Liu2021VariationCA}
\bibfield{author}{\bibinfo{person}{Shuchang Liu}, \bibinfo{person}{Fei Sun}, \bibinfo{person}{Yingqiang Ge}, \bibinfo{person}{Changhua Pei}, {and} \bibinfo{person}{Yongfeng Zhang}.} \bibinfo{year}{2021}\natexlab{}.
\newblock \showarticletitle{Variation Control and Evaluation for Generative Slate Recommendations}.
\newblock \bibinfo{journal}{\emph{Proceedings of the Web Conference 2021}} (\bibinfo{year}{2021}).
\newblock
\urldef\tempurl%
\url{https://api.semanticscholar.org/CorpusID:232069075}
\showURL{%
\tempurl}


\bibitem[Liu et~al\mbox{.}(2022)]%
        {liu2022neural}
\bibfield{author}{\bibinfo{person}{Weiwen Liu}, \bibinfo{person}{Yunjia Xi}, \bibinfo{person}{Jiarui Qin}, \bibinfo{person}{Fei Sun}, \bibinfo{person}{Bo Chen}, \bibinfo{person}{Weinan Zhang}, \bibinfo{person}{Rui Zhang}, {and} \bibinfo{person}{Ruiming Tang}.} \bibinfo{year}{2022}\natexlab{}.
\newblock \showarticletitle{Neural Re-ranking in Multi-stage Recommender Systems: A Review}.
\newblock \bibinfo{journal}{\emph{arXiv preprint arXiv:2202.06602}} (\bibinfo{year}{2022}).
\newblock


\bibitem[Lops et~al\mbox{.}(2023)]%
        {lops2023clayrs}
\bibfield{author}{\bibinfo{person}{Pasquale Lops}, \bibinfo{person}{Marco Polignano}, \bibinfo{person}{Cataldo Musto}, \bibinfo{person}{Antonio Silletti}, {and} \bibinfo{person}{Giovanni Semeraro}.} \bibinfo{year}{2023}\natexlab{}.
\newblock \showarticletitle{ClayRS: An end-to-end framework for reproducible knowledge-aware recommender systems}.
\newblock \bibinfo{journal}{\emph{Information Systems}}  \bibinfo{volume}{119} (\bibinfo{year}{2023}), \bibinfo{pages}{102273}.
\newblock


\bibitem[Lv et~al\mbox{.}(2023)]%
        {lv2023deep}
\bibfield{author}{\bibinfo{person}{Yimin Lv}, \bibinfo{person}{Shuli Wang}, \bibinfo{person}{Beihong Jin}, \bibinfo{person}{Yisong Yu}, \bibinfo{person}{Yapeng Zhang}, \bibinfo{person}{Jian Dong}, \bibinfo{person}{Yongkang Wang}, \bibinfo{person}{Xingxing Wang}, {and} \bibinfo{person}{Dong Wang}.} \bibinfo{year}{2023}\natexlab{}.
\newblock \showarticletitle{Deep Situation-Aware Interaction Network for Click-Through Rate Prediction}. In \bibinfo{booktitle}{\emph{Proceedings of the 17th ACM Conference on Recommender Systems}}. \bibinfo{pages}{171--182}.
\newblock


\bibitem[Mao et~al\mbox{.}(2023)]%
        {mao2023finalmlp}
\bibfield{author}{\bibinfo{person}{Kelong Mao}, \bibinfo{person}{Jieming Zhu}, \bibinfo{person}{Liangcai Su}, \bibinfo{person}{Guohao Cai}, \bibinfo{person}{Yuru Li}, {and} \bibinfo{person}{Zhenhua Dong}.} \bibinfo{year}{2023}\natexlab{}.
\newblock \showarticletitle{FinalMLP: an enhanced two-stream MLP model for CTR prediction}. In \bibinfo{booktitle}{\emph{Proceedings of the AAAI Conference on Artificial Intelligence}}, Vol.~\bibinfo{volume}{37}. \bibinfo{pages}{4552--4560}.
\newblock


\bibitem[Michiels et~al\mbox{.}(2022)]%
        {lien2022RecPack}
\bibfield{author}{\bibinfo{person}{Lien Michiels}, \bibinfo{person}{Robin Verachtert}, {and} \bibinfo{person}{Bart Goethals}.} \bibinfo{year}{2022}\natexlab{}.
\newblock \showarticletitle{RecPack: An(Other) Experimentation Toolkit for Top-N Recommendation Using Implicit Feedback Data}. In \bibinfo{booktitle}{\emph{Proceedings of the 16th ACM Conference on Recommender Systems}} (Seattle, WA, USA) \emph{(\bibinfo{series}{RecSys '22})}. \bibinfo{publisher}{Association for Computing Machinery}, \bibinfo{address}{New York, NY, USA}, \bibinfo{pages}{648–651}.
\newblock
\showISBNx{9781450392785}
\urldef\tempurl%
\url{https://doi.org/10.1145/3523227.3551472}
\showDOI{\tempurl}


\bibitem[Pang et~al\mbox{.}(2020)]%
        {pang2020setrank}
\bibfield{author}{\bibinfo{person}{Liang Pang}, \bibinfo{person}{Jun Xu}, \bibinfo{person}{Qingyao Ai}, \bibinfo{person}{Yanyan Lan}, \bibinfo{person}{Xueqi Cheng}, {and} \bibinfo{person}{Jirong Wen}.} \bibinfo{year}{2020}\natexlab{}.
\newblock \showarticletitle{Setrank: Learning a permutation-invariant ranking model for information retrieval}. In \bibinfo{booktitle}{\emph{Proceedings of the 43rd international ACM SIGIR conference on research and development in information retrieval}}. \bibinfo{pages}{499--508}.
\newblock


\bibitem[Pei et~al\mbox{.}(2019)]%
        {pei2019PRM}
\bibfield{author}{\bibinfo{person}{Changhua Pei}, \bibinfo{person}{Yi Zhang}, \bibinfo{person}{Yongfeng Zhang}, \bibinfo{person}{Fei Sun}, \bibinfo{person}{Xiao Lin}, \bibinfo{person}{Hanxiao Sun}, \bibinfo{person}{Jian Wu}, \bibinfo{person}{Peng Jiang}, \bibinfo{person}{Junfeng Ge}, \bibinfo{person}{Wenwu Ou}, {et~al\mbox{.}}} \bibinfo{year}{2019}\natexlab{}.
\newblock \showarticletitle{Personalized re-ranking for recommendation}. In \bibinfo{booktitle}{\emph{Proceedings of the 13th ACM conference on recommender systems}}. \bibinfo{pages}{3--11}.
\newblock


\bibitem[Ren et~al\mbox{.}(2023)]%
        {Ren2023SlateAwareRF}
\bibfield{author}{\bibinfo{person}{Yi Ren}, \bibinfo{person}{Xiao Han}, \bibinfo{person}{Xu-Hui Zhao}, \bibinfo{person}{Shenzheng Zhang}, {and} \bibinfo{person}{Yan Zhang}.} \bibinfo{year}{2023}\natexlab{}.
\newblock \showarticletitle{Slate-Aware Ranking for Recommendation}.
\newblock \bibinfo{journal}{\emph{Proceedings of the Sixteenth ACM International Conference on Web Search and Data Mining}} (\bibinfo{year}{2023}).
\newblock
\urldef\tempurl%
\url{https://api.semanticscholar.org/CorpusID:257079671}
\showURL{%
\tempurl}


\bibitem[Rendle(2010)]%
        {rendle2010fm}
\bibfield{author}{\bibinfo{person}{Steffen Rendle}.} \bibinfo{year}{2010}\natexlab{}.
\newblock \showarticletitle{Factorization machines}. In \bibinfo{booktitle}{\emph{2010 IEEE International conference on data mining}}. IEEE, \bibinfo{pages}{995--1000}.
\newblock


\bibitem[Rendle et~al\mbox{.}(2012)]%
        {rendle2012bpr}
\bibfield{author}{\bibinfo{person}{Steffen Rendle}, \bibinfo{person}{Christoph Freudenthaler}, \bibinfo{person}{Zeno Gantner}, {and} \bibinfo{person}{Lars Schmidt-Thieme}.} \bibinfo{year}{2012}\natexlab{}.
\newblock \showarticletitle{BPR: Bayesian personalized ranking from implicit feedback}.
\newblock \bibinfo{journal}{\emph{arXiv preprint arXiv:1205.2618}} (\bibinfo{year}{2012}).
\newblock


\bibitem[Rendle et~al\mbox{.}(2010)]%
        {rendle2010fpmc}
\bibfield{author}{\bibinfo{person}{Steffen Rendle}, \bibinfo{person}{Christoph Freudenthaler}, {and} \bibinfo{person}{Lars Schmidt-Thieme}.} \bibinfo{year}{2010}\natexlab{}.
\newblock \showarticletitle{Factorizing personalized markov chains for next-basket recommendation}. In \bibinfo{booktitle}{\emph{Proceedings of the 19th international conference on World wide web}}. \bibinfo{pages}{811--820}.
\newblock


\bibitem[Salah et~al\mbox{.}(2020)]%
        {salah2020cornac}
\bibfield{author}{\bibinfo{person}{Aghiles Salah}, \bibinfo{person}{Quoc-Tuan Truong}, {and} \bibinfo{person}{Hady~W Lauw}.} \bibinfo{year}{2020}\natexlab{}.
\newblock \showarticletitle{Cornac: A Comparative Framework for Multimodal Recommender Systems}.
\newblock \bibinfo{journal}{\emph{Journal of Machine Learning Research}} \bibinfo{volume}{21}, \bibinfo{number}{95} (\bibinfo{year}{2020}), \bibinfo{pages}{1--5}.
\newblock


\bibitem[Song et~al\mbox{.}(2019)]%
        {song2019autoint}
\bibfield{author}{\bibinfo{person}{Weiping Song}, \bibinfo{person}{Chence Shi}, \bibinfo{person}{Zhiping Xiao}, \bibinfo{person}{Zhijian Duan}, \bibinfo{person}{Yewen Xu}, \bibinfo{person}{Ming Zhang}, {and} \bibinfo{person}{Jian Tang}.} \bibinfo{year}{2019}\natexlab{}.
\newblock \showarticletitle{Autoint: Automatic feature interaction learning via self-attentive neural networks}. In \bibinfo{booktitle}{\emph{Proceedings of the 28th ACM international conference on information and knowledge management}}. \bibinfo{pages}{1161--1170}.
\newblock


\bibitem[Sun(2023a)]%
        {sun2023challenges}
\bibfield{author}{\bibinfo{person}{Aixin Sun}.} \bibinfo{year}{2023}\natexlab{a}.
\newblock \showarticletitle{On Challenges of Evaluating Recommender Systems in an Offline Setting}. In \bibinfo{booktitle}{\emph{Proceedings of the 17th ACM Conference on Recommender Systems}}. \bibinfo{pages}{1284--1285}.
\newblock


\bibitem[Sun(2023b)]%
        {sun2023take}
\bibfield{author}{\bibinfo{person}{Aixin Sun}.} \bibinfo{year}{2023}\natexlab{b}.
\newblock \showarticletitle{Take a Fresh Look at Recommender Systems from an Evaluation Standpoint}. In \bibinfo{booktitle}{\emph{Proceedings of the 46th International ACM SIGIR Conference on Research and Development in Information Retrieval}}. \bibinfo{pages}{2629--2638}.
\newblock


\bibitem[Sun et~al\mbox{.}(2023)]%
        {sun2023neighborhood}
\bibfield{author}{\bibinfo{person}{Peijie Sun}, \bibinfo{person}{Le Wu}, \bibinfo{person}{Kun Zhang}, \bibinfo{person}{Xiangzhi Chen}, {and} \bibinfo{person}{Meng Wang}.} \bibinfo{year}{2023}\natexlab{}.
\newblock \showarticletitle{Neighborhood-Enhanced Supervised Contrastive Learning for Collaborative Filtering}.
\newblock \bibinfo{journal}{\emph{IEEE Transactions on Knowledge and Data Engineering}} (\bibinfo{year}{2023}).
\newblock


\bibitem[Sun et~al\mbox{.}(2022)]%
        {sun2022daisyrec}
\bibfield{author}{\bibinfo{person}{Zhu Sun}, \bibinfo{person}{Hui Fang}, \bibinfo{person}{Jie Yang}, \bibinfo{person}{Xinghua Qu}, \bibinfo{person}{Hongyang Liu}, \bibinfo{person}{Di Yu}, \bibinfo{person}{Yew-Soon Ong}, {and} \bibinfo{person}{Jie Zhang}.} \bibinfo{year}{2022}\natexlab{}.
\newblock \showarticletitle{DaisyRec 2.0: Benchmarking Recommendation for Rigorous Evaluation}.
\newblock \bibinfo{journal}{\emph{IEEE Transactions on Pattern Analysis and Machine Intelligence (TPAMI)}} (\bibinfo{year}{2022}).
\newblock


\bibitem[Sun et~al\mbox{.}(2020)]%
        {sun2020we}
\bibfield{author}{\bibinfo{person}{Zhu Sun}, \bibinfo{person}{Di Yu}, \bibinfo{person}{Hui Fang}, \bibinfo{person}{Jie Yang}, \bibinfo{person}{Xinghua Qu}, \bibinfo{person}{Jie Zhang}, {and} \bibinfo{person}{Cong Geng}.} \bibinfo{year}{2020}\natexlab{}.
\newblock \showarticletitle{Are we evaluating rigorously? benchmarking recommendation for reproducible evaluation and fair comparison}. In \bibinfo{booktitle}{\emph{Proceedings of the 14th ACM Conference on Recommender Systems}}. \bibinfo{pages}{23--32}.
\newblock


\bibitem[Tang and Wang(2018)]%
        {tang2018caser}
\bibfield{author}{\bibinfo{person}{Jiaxi Tang} {and} \bibinfo{person}{Ke Wang}.} \bibinfo{year}{2018}\natexlab{}.
\newblock \showarticletitle{Personalized top-n sequential recommendation via convolutional sequence embedding}. In \bibinfo{booktitle}{\emph{Proceedings of the eleventh ACM international conference on web search and data mining}}. \bibinfo{pages}{565--573}.
\newblock


\bibitem[Wang et~al\mbox{.}(2023a)]%
        {wang2023contrarec}
\bibfield{author}{\bibinfo{person}{Chenyang Wang}, \bibinfo{person}{Weizhi Ma}, \bibinfo{person}{Chong Chen}, \bibinfo{person}{Min Zhang}, \bibinfo{person}{Yiqun Liu}, {and} \bibinfo{person}{Shaoping Ma}.} \bibinfo{year}{2023}\natexlab{a}.
\newblock \showarticletitle{Sequential recommendation with multiple contrast signals}.
\newblock \bibinfo{journal}{\emph{ACM Transactions on Information Systems}} \bibinfo{volume}{41}, \bibinfo{number}{1} (\bibinfo{year}{2023}), \bibinfo{pages}{1--27}.
\newblock


\bibitem[Wang et~al\mbox{.}(2020a)]%
        {wang2020kda}
\bibfield{author}{\bibinfo{person}{Chenyang Wang}, \bibinfo{person}{Weizhi Ma}, \bibinfo{person}{Min Zhang}, \bibinfo{person}{Chong Chen}, \bibinfo{person}{Yiqun Liu}, {and} \bibinfo{person}{Shaoping Ma}.} \bibinfo{year}{2020}\natexlab{a}.
\newblock \showarticletitle{Toward dynamic user intention: Temporal evolutionary effects of item relations in sequential recommendation}.
\newblock \bibinfo{journal}{\emph{ACM Transactions on Information Systems (TOIS)}} \bibinfo{volume}{39}, \bibinfo{number}{2} (\bibinfo{year}{2020}), \bibinfo{pages}{1--33}.
\newblock


\bibitem[Wang et~al\mbox{.}(2022a)]%
        {wang2022timirec}
\bibfield{author}{\bibinfo{person}{Chenyang Wang}, \bibinfo{person}{Zhefan Wang}, \bibinfo{person}{Yankai Liu}, \bibinfo{person}{Yang Ge}, \bibinfo{person}{Weizhi Ma}, \bibinfo{person}{Min Zhang}, \bibinfo{person}{Yiqun Liu}, \bibinfo{person}{Junlan Feng}, \bibinfo{person}{Chao Deng}, {and} \bibinfo{person}{Shaoping Ma}.} \bibinfo{year}{2022}\natexlab{a}.
\newblock \showarticletitle{Target interest distillation for multi-interest recommendation}. In \bibinfo{booktitle}{\emph{Proceedings of the 31st ACM International Conference on Information \& Knowledge Management}}. \bibinfo{pages}{2007--2016}.
\newblock


\bibitem[Wang et~al\mbox{.}(2022b)]%
        {wang2022directau}
\bibfield{author}{\bibinfo{person}{Chenyang Wang}, \bibinfo{person}{Yuanqing Yu}, \bibinfo{person}{Weizhi Ma}, \bibinfo{person}{Min Zhang}, \bibinfo{person}{Chong Chen}, \bibinfo{person}{Yiqun Liu}, {and} \bibinfo{person}{Shaoping Ma}.} \bibinfo{year}{2022}\natexlab{b}.
\newblock \showarticletitle{Towards representation alignment and uniformity in collaborative filtering}. In \bibinfo{booktitle}{\emph{Proceedings of the 28th ACM SIGKDD conference on knowledge discovery and data mining}}. \bibinfo{pages}{1816--1825}.
\newblock


\bibitem[Wang et~al\mbox{.}(2019)]%
        {wang2019SLRCPlus}
\bibfield{author}{\bibinfo{person}{Chenyang Wang}, \bibinfo{person}{Min Zhang}, \bibinfo{person}{Weizhi Ma}, \bibinfo{person}{Yiqun Liu}, {and} \bibinfo{person}{Shaoping Ma}.} \bibinfo{year}{2019}\natexlab{}.
\newblock \showarticletitle{Modeling item-specific temporal dynamics of repeat consumption for recommender systems}. In \bibinfo{booktitle}{\emph{The world wide web conference}}. \bibinfo{pages}{1977--1987}.
\newblock


\bibitem[Wang et~al\mbox{.}(2020b)]%
        {wang2020make}
\bibfield{author}{\bibinfo{person}{Chenyang Wang}, \bibinfo{person}{Min Zhang}, \bibinfo{person}{Weizhi Ma}, \bibinfo{person}{Yiqun Liu}, {and} \bibinfo{person}{Shaoping Ma}.} \bibinfo{year}{2020}\natexlab{b}.
\newblock \showarticletitle{Make it a chorus: knowledge-and time-aware item modeling for sequential recommendation}. In \bibinfo{booktitle}{\emph{Proceedings of the 43rd International ACM SIGIR Conference on Research and Development in Information Retrieval}}. \bibinfo{pages}{109--118}.
\newblock


\bibitem[Wang et~al\mbox{.}(2017)]%
        {wang2017dcn}
\bibfield{author}{\bibinfo{person}{Ruoxi Wang}, \bibinfo{person}{Bin Fu}, \bibinfo{person}{Gang Fu}, {and} \bibinfo{person}{Mingliang Wang}.} \bibinfo{year}{2017}\natexlab{}.
\newblock \showarticletitle{Deep \& cross network for ad click predictions}.
\newblock In \bibinfo{booktitle}{\emph{Proceedings of the ADKDD'17}}. \bibinfo{pages}{1--7}.
\newblock


\bibitem[Wang et~al\mbox{.}(2021)]%
        {wang2021dcnv2}
\bibfield{author}{\bibinfo{person}{Ruoxi Wang}, \bibinfo{person}{Rakesh Shivanna}, \bibinfo{person}{Derek Cheng}, \bibinfo{person}{Sagar Jain}, \bibinfo{person}{Dong Lin}, \bibinfo{person}{Lichan Hong}, {and} \bibinfo{person}{Ed Chi}.} \bibinfo{year}{2021}\natexlab{}.
\newblock \showarticletitle{Dcn v2: Improved deep \& cross network and practical lessons for web-scale learning to rank systems}. In \bibinfo{booktitle}{\emph{Proceedings of the web conference 2021}}. \bibinfo{pages}{1785--1797}.
\newblock


\bibitem[Wang et~al\mbox{.}(2023b)]%
        {wang2023unbiased}
\bibfield{author}{\bibinfo{person}{Yifan Wang}, \bibinfo{person}{Peijie Sun}, \bibinfo{person}{Min Zhang}, \bibinfo{person}{Qinglin Jia}, \bibinfo{person}{Jingjie Li}, {and} \bibinfo{person}{Shaoping Ma}.} \bibinfo{year}{2023}\natexlab{b}.
\newblock \showarticletitle{Unbiased Delayed Feedback Label Correction for Conversion Rate Prediction}. In \bibinfo{booktitle}{\emph{Proceedings of the 29th ACM SIGKDD Conference on Knowledge Discovery and Data Mining}}. \bibinfo{pages}{2456--2466}.
\newblock


\bibitem[Wu et~al\mbox{.}(2020)]%
        {wu2020mind}
\bibfield{author}{\bibinfo{person}{Fangzhao Wu}, \bibinfo{person}{Ying Qiao}, \bibinfo{person}{Jiun-Hung Chen}, \bibinfo{person}{Chuhan Wu}, \bibinfo{person}{Tao Qi}, \bibinfo{person}{Jianxun Lian}, \bibinfo{person}{Danyang Liu}, \bibinfo{person}{Xing Xie}, \bibinfo{person}{Jianfeng Gao}, \bibinfo{person}{Winnie Wu}, {et~al\mbox{.}}} \bibinfo{year}{2020}\natexlab{}.
\newblock \showarticletitle{Mind: A large-scale dataset for news recommendation}. In \bibinfo{booktitle}{\emph{Proceedings of the 58th annual meeting of the association for computational linguistics}}. \bibinfo{pages}{3597--3606}.
\newblock


\bibitem[Xi et~al\mbox{.}(2022a)]%
        {xi_context-aware_2022}
\bibfield{author}{\bibinfo{person}{Yunjia Xi}, \bibinfo{person}{Weiwen Liu}, \bibinfo{person}{Xinyi Dai}, \bibinfo{person}{Ruiming Tang}, \bibinfo{person}{Weinan Zhang}, \bibinfo{person}{Qing Liu}, \bibinfo{person}{Xiuqiang He}, {and} \bibinfo{person}{Yong Yu}.} \bibinfo{year}{2022}\natexlab{a}.
\newblock \showarticletitle{Context-aware {Reranking} with {Utility} {Maximization} for {Recommendation}}.
\newblock \bibinfo{journal}{\emph{arXiv}} (\bibinfo{date}{Feb.} \bibinfo{year}{2022}).
\newblock


\bibitem[Xi et~al\mbox{.}(2022b)]%
        {xi_multi-level_2022}
\bibfield{author}{\bibinfo{person}{Yunjia Xi}, \bibinfo{person}{Weiwen Liu}, \bibinfo{person}{Jieming Zhu}, \bibinfo{person}{Xilong Zhao}, \bibinfo{person}{Xinyi Dai}, \bibinfo{person}{Ruiming Tang}, \bibinfo{person}{Weinan Zhang}, \bibinfo{person}{Rui Zhang}, {and} \bibinfo{person}{Yong Yu}.} \bibinfo{year}{2022}\natexlab{b}.
\newblock \bibinfo{title}{Multi-{Level} {Interaction} {Reranking} with {User} {Behavior} {History}}.
\newblock
\newblock


\bibitem[Xiao et~al\mbox{.}(2017)]%
        {xiao2017afm}
\bibfield{author}{\bibinfo{person}{Jun Xiao}, \bibinfo{person}{Hao Ye}, \bibinfo{person}{Xiangnan He}, \bibinfo{person}{Hanwang Zhang}, \bibinfo{person}{Fei Wu}, {and} \bibinfo{person}{Tat-Seng Chua}.} \bibinfo{year}{2017}\natexlab{}.
\newblock \showarticletitle{Attentional factorization machines: Learning the weight of feature interactions via attention networks}.
\newblock \bibinfo{journal}{\emph{arXiv preprint arXiv:1708.04617}} (\bibinfo{year}{2017}).
\newblock


\bibitem[Yang and Zhai(2022)]%
        {yang2022click}
\bibfield{author}{\bibinfo{person}{Yanwu Yang} {and} \bibinfo{person}{Panyu Zhai}.} \bibinfo{year}{2022}\natexlab{}.
\newblock \showarticletitle{Click-through rate prediction in online advertising: A literature review}.
\newblock \bibinfo{journal}{\emph{Information Processing \& Management}} \bibinfo{volume}{59}, \bibinfo{number}{2} (\bibinfo{year}{2022}), \bibinfo{pages}{102853}.
\newblock


\bibitem[Yu et~al\mbox{.}(2024)]%
        {yu2024easyrl4rec}
\bibfield{author}{\bibinfo{person}{Yuanqing Yu}, \bibinfo{person}{Chongming Gao}, \bibinfo{person}{Jiawei Chen}, \bibinfo{person}{Heng Tang}, \bibinfo{person}{Yuefeng Sun}, \bibinfo{person}{Qian Chen}, \bibinfo{person}{Weizhi Ma}, {and} \bibinfo{person}{Min Zhang}.} \bibinfo{year}{2024}\natexlab{}.
\newblock \showarticletitle{EasyRL4Rec: A User-Friendly Code Library for Reinforcement Learning Based Recommender Systems}.
\newblock \bibinfo{journal}{\emph{arXiv preprint arXiv:2402.15164}} (\bibinfo{year}{2024}).
\newblock


\bibitem[Zhang et~al\mbox{.}(2018)]%
        {zhang2018cfkg}
\bibfield{author}{\bibinfo{person}{Yongfeng Zhang}, \bibinfo{person}{Qingyao Ai}, \bibinfo{person}{Xu Chen}, {and} \bibinfo{person}{Pengfei Wang}.} \bibinfo{year}{2018}\natexlab{}.
\newblock \showarticletitle{Learning over knowledge-base embeddings for recommendation}.
\newblock \bibinfo{journal}{\emph{arXiv preprint arXiv:1803.06540}} (\bibinfo{year}{2018}).
\newblock


\bibitem[Zhao et~al\mbox{.}(2022)]%
        {zhao2022recbole[2.0]}
\bibfield{author}{\bibinfo{person}{Wayne~Xin Zhao}, \bibinfo{person}{Yupeng Hou}, \bibinfo{person}{Xingyu Pan}, \bibinfo{person}{Chen Yang}, \bibinfo{person}{Zeyu Zhang}, \bibinfo{person}{Zihan Lin}, \bibinfo{person}{Jingsen Zhang}, \bibinfo{person}{Shuqing Bian}, \bibinfo{person}{Jiakai Tang}, \bibinfo{person}{Wenqi Sun}, \bibinfo{person}{Yushuo Chen}, \bibinfo{person}{Lanling Xu}, \bibinfo{person}{Gaowei Zhang}, \bibinfo{person}{Zhen Tian}, \bibinfo{person}{Changxin Tian}, \bibinfo{person}{Shanlei Mu}, \bibinfo{person}{Xinyan Fan}, \bibinfo{person}{Xu Chen}, {and} \bibinfo{person}{Ji{-}Rong Wen}.} \bibinfo{year}{2022}\natexlab{}.
\newblock \showarticletitle{RecBole 2.0: Towards a More Up-to-Date Recommendation Library}. In \bibinfo{booktitle}{\emph{{CIKM}}}. \bibinfo{publisher}{{ACM}}, \bibinfo{pages}{4722--4726}.
\newblock


\bibitem[Zhou et~al\mbox{.}(2019)]%
        {zhou2019dien}
\bibfield{author}{\bibinfo{person}{Guorui Zhou}, \bibinfo{person}{Na Mou}, \bibinfo{person}{Ying Fan}, \bibinfo{person}{Qi Pi}, \bibinfo{person}{Weijie Bian}, \bibinfo{person}{Chang Zhou}, \bibinfo{person}{Xiaoqiang Zhu}, {and} \bibinfo{person}{Kun Gai}.} \bibinfo{year}{2019}\natexlab{}.
\newblock \showarticletitle{Deep interest evolution network for click-through rate prediction}. In \bibinfo{booktitle}{\emph{Proceedings of the AAAI conference on artificial intelligence}}, Vol.~\bibinfo{volume}{33}. \bibinfo{pages}{5941--5948}.
\newblock


\bibitem[Zhou et~al\mbox{.}(2018)]%
        {zhou2018din}
\bibfield{author}{\bibinfo{person}{Guorui Zhou}, \bibinfo{person}{Xiaoqiang Zhu}, \bibinfo{person}{Chenru Song}, \bibinfo{person}{Ying Fan}, \bibinfo{person}{Han Zhu}, \bibinfo{person}{Xiao Ma}, \bibinfo{person}{Yanghui Yan}, \bibinfo{person}{Junqi Jin}, \bibinfo{person}{Han Li}, {and} \bibinfo{person}{Kun Gai}.} \bibinfo{year}{2018}\natexlab{}.
\newblock \showarticletitle{Deep interest network for click-through rate prediction}. In \bibinfo{booktitle}{\emph{Proceedings of the 24th ACM SIGKDD international conference on knowledge discovery \& data mining}}. \bibinfo{pages}{1059--1068}.
\newblock


\bibitem[Zhu et~al\mbox{.}(2022)]%
        {zhu2022bars}
\bibfield{author}{\bibinfo{person}{Jieming Zhu}, \bibinfo{person}{Quanyu Dai}, \bibinfo{person}{Liangcai Su}, \bibinfo{person}{Rong Ma}, \bibinfo{person}{Jinyang Liu}, \bibinfo{person}{Guohao Cai}, \bibinfo{person}{Xi Xiao}, {and} \bibinfo{person}{Rui Zhang}.} \bibinfo{year}{2022}\natexlab{}.
\newblock \showarticletitle{{BARS:} Towards Open Benchmarking for Recommender Systems}. In \bibinfo{booktitle}{\emph{{SIGIR} '22: The 45th International {ACM} {SIGIR} Conference on Research and Development in Information Retrieval, Madrid, Spain, July 11 - 15, 2022}}, \bibfield{editor}{\bibinfo{person}{Enrique Amig{\'{o}}}, \bibinfo{person}{Pablo Castells}, \bibinfo{person}{Julio Gonzalo}, \bibinfo{person}{Ben Carterette}, \bibinfo{person}{J.~Shane Culpepper}, {and} \bibinfo{person}{Gabriella Kazai}} (Eds.). \bibinfo{publisher}{{ACM}}, \bibinfo{pages}{2912--2923}.
\newblock
\urldef\tempurl%
\url{https://doi.org/10.1145/3477495.3531723}
\showDOI{\tempurl}


\end{thebibliography}
\end{document}